\documentclass[graybox]{svmult}

\usepackage{mathptmx}       
\usepackage{helvet}         
\usepackage{courier}        
\usepackage{type1cm}        
\usepackage{makeidx}         
\usepackage{graphicx}        
\usepackage{multicol}        
\usepackage[bottom]{footmisc}

\usepackage{amsmath,amssymb,braket,bbold,esvect}

\UseRawInputEncoding

\makeindex             

\begin{document}

\title*{Topological Crystalline Insulators}
\author{Titus Neupert and Frank Schindler}
\institute{Titus Neupert \at Department of Physics, University of Zurich,\\ Wintherthurerstrasse 190, CH-8057 Zurich, Switzerland, \email{titus.neupert@uzh.ch}
\and Frank Schindler \at Department of Physics, University of Zurich,\\ Wintherthurerstrasse 190, CH-8057 Zurich, Switzerland, \email{frank.schindler@uzh.ch}}
%
%
\maketitle

\abstract{We give an introduction to topological crystalline insulators, that is, gapped ground states of quantum matter that 
are not adiabatically connected to an atomic limit without breaking symmetries that include spatial transformations, like mirror or rotational symmetries. 
To deduce the topological properties, we use non-Abelian Wilson loops.
We also discuss in detail higher-order topological insulators with hinge and corner states, and in particular present interacting bosonic models for the latter class of systems.
}

\newpage
\section{Wilson loops and the bulk-boundary correspondence}
We first provide a unified picture of topological bulk-boundary correspondences in any dimension by making use of Brillouin zone Wilson loops.
\subsection{Introduction and motivation}
\label{sec: motivation}
In these notes, we are mostly concerned with the topological characterization of non-interacting electron Hamiltonians on a lattice in the presence of spatial symmetries. In general, an insulating topological phase of matter may be defined by the requirement that the many-body ground state of the corresponding Hamiltonian (given by a Slater determinant in the non-interacting case) cannot be adiabatically connected to the atomic limit of vanishing hopping between the sites of the lattice. Further requiring that certain symmetries such as time-reversal are not violated along any such adiabatic interpolation enriches the topological classification, in that phases which were classified as trivial in the previous sense now acquire a topological distinction which is protected by the respective symmetry.

To determine the topology of a given ground state, several topological invariants have been proposed, such as the Pfaffian invariant for two-dimensional~(2D) time-reversal symmetric systems. However, they often require Bloch states to be provided in a smooth gauge across the whole Brillouin zone~(BZ) for their evaluation, making them impractical for numerical calculations. In addition, most of them are specific to the dimension or symmetry class considered and thus do not generalize well.

Here, we employ non-Abelian Wilson loops as a generalization of the one-dimensional~(1D) Berry phase to characterize topological properties in any dimension and any symmetry class. This provides a framework of topological invariants which makes direct contact with the protected boundary degrees of freedom of a given phase.

As a prerequisite, we assume a working knowledge of the (boundary) physics of non-crystalline topological phases and their topological invariants, as well as their classification by the tenfold way. Suitable introductions can be found in Refs.~\cite{Qi08,HasanKaneColloq,BernevigBook,AsbothIntro,BernevigNeupertLectures}.


\subsection{Definitions}
We work in units where $\hslash = c = e = 1$ and denote by $\sigma_i$, $i=x,y,z$, the $2 \times 2$ Pauli matrices. We define $\sigma_0 = \mathbb{1}_{2 \times 2}$ for convenience. We express eigenfunctions of a translationally invariant single-particle Hamiltonian in the basis
\begin{equation}
\phi_{\mathbf{k},\alpha}(\mathbf{r})=\frac{1}{\sqrt{N}}
\sum_{\mathbf{R}}e^{\mathrm{i}\mathbf{k}\cdot(\mathbf{R}+\mathbf{r}_\alpha)}
\varphi_{\mathbf{R},\alpha}(\mathbf{r}-\mathbf{R}-\mathbf{r}_\alpha),
\end{equation}
where $\varphi_{\mathbf{R},\alpha}$, $\alpha=1,\cdots, N$, are the orbitals chosen as basis for the finite-dimensional Hilbert space in each unit cell, labelled by the lattice vector $\mathbf{R}$, and $\mathbf{r}_\alpha$ is the center of each of these orbitals relative to the origin of the unit cell. 
Including $\mathbf{r}_\alpha$ in the exponential corresponds to a convenient choice of gauge when studying the response to external fields defined in continuous real space. 

A general non-interacting Hamiltonian then has the Bloch matrix elements
\begin{equation}
\mathcal{H}_{\alpha,\beta}(\mathbf{k})=
\int \mathrm{d}^d r
\phi^*_{\mathbf{k},\alpha}(\mathbf{r})\hat{H}\phi_{\mathbf{k},\beta}(\mathbf{r}),
\end{equation}
as well as energy eigenstates
\begin{equation} \label{eq: fullEigenstates}
\psi_{\mathbf{k},n}(\mathbf{r})=\sum_\alpha^N\, u_{\mathbf{k};n,\alpha}\phi_{\mathbf{k},\alpha}(\mathbf{r}),
\end{equation}
where
\begin{equation}
\sum_\beta\mathcal{H}_{\alpha,\beta}(\mathbf{k}) u_{\mathbf{k};n,\beta}= \epsilon_{n}(\mathbf{k})  u_{\mathbf{k};n,\alpha}, \qquad n=1,\cdots, N.
\end{equation}
In the following, we are interested in situations where the system has an energy gap after the first $M<N$ bands, i.e., $\epsilon_{M}(\mathbf{k})<\epsilon_{M+1}(\mathbf{k})$ for all $\mathbf{k}$.

\subsection{Wilson loop and position operator}
\label{sec: WilsonLoopPositionOperator}
Introduced in 1984 by Sir Michael Berry, the so-called Berry phase describes a phase factor which arises in addition to the dynamical evolution $e^{\mathrm{i} \int E[\lambda(t)] dt}$ of a quantum mechanical state in an adiabatic interpolation of the corresponding Hamiltonian $\hat{H}[\lambda(t)]$ along a closed path $\lambda(t)$ in parameter space. It depends only on the geometry of the path chosen, and can be expressed as a line integral of the Berry connection, which we define below for the case where the parameter $\lambda$ is a single particle momentum. If degeneracies between energy levels are encountered along the path, we have to consider the joint evolution of a set of eigenstates that may have degeneracies. If we consider $M$ such states, the Berry phase generalizes to a $U(M)$ matrix, which may be expressed as the line integral of a non-Abelian Berry connection, and is called non-Abelian Wilson loop.

In the BZ, we may consider momentum $\mathbf{k}$ as a parameter of the Bloch Hamiltonian $\mathcal{H}(\mathbf{k})$. The corresponding non-Abelian Berry-Wilczek-Zee connection is then given by
\begin{equation}
\mathbf{A}_{m,n}(\mathbf{k})=\langle u_{\mathbf{k},m}|\mathbf{\nabla}_{\mathbf{k}}|u_{\mathbf{k},n}\rangle,\qquad n,m=1,\cdots,M.
\end{equation}
Note that it is anti-Hermitian, that is, it satisfies $\mathbf{A}^*_{n,m}(\mathbf{k}) = - \mathbf{A}_{m,n}(\mathbf{k})$.
Using matrix notation, we define the Wilson loop, a unitary operator, as
\begin{equation} \label{eq: WilsonLoop}
W[l]=\overline{\mathrm{exp}}\left[-\int_l \mathrm{d}\mathbf{l}\cdot \mathbf{A}(\mathbf{k})\right],
\end{equation}
where $l$ is a loop in momentum space and the overline denotes path ordering of the exponential, where as usual operators at the beginning of the path occur to the right of operators at the end. This unitary operator acts on the occupied band manifold, and can be numerically evaluated with the formula
\begin{eqnarray} \label{eq: WLNumericalFormula}
W_{n_{R+1},n_1}[l]  
&=& \, 
\lim_{R\to \infty}
\sum_{n_2,\cdots n_{R}=1}^M
\prod_{i=R}^1
\Bigl[ \mathrm{exp}\left[- (\mathbf{k}_{i+1}-\mathbf{k}_{i})\cdot \mathbf{A}(\mathbf{k}_{i+1})\right] \Bigr]_{n_{i+1},n_i}
\nonumber\\
&=& \, 
\lim_{R\to \infty}
\sum_{n_2,\cdots n_{R}=1}^M
\prod_{i=R}^1
\Bigl[\delta_{n_{i+1},n_{i}}- (\mathbf{k}_{i+1}-\mathbf{k}_{i})\cdot \mathbf{A}_{n_{i+1},n_i}(\mathbf{k}_{i+1}) \Bigr]
\nonumber\\
&=& \, 
\lim_{R\to \infty}
\sum_{n_2,\cdots n_{R}=1}^M
\prod_{i=R}^1
\Bigl[\braket{u_{\mathbf{k}_{i+1},n_{i+1}}|u_{\mathbf{k}_{i+1},n_{i}}} \\&&- (\mathbf{k}_{i+1}-\mathbf{k}_{i})\cdot 
\braket{u_{\mathbf{k}_{i+1},n_{i+1}}|\mathbf{\nabla}_{\mathbf{k}_{i+1}}|u_{\mathbf{k}_{i+1},n_{i}}}
 \Bigr]
\nonumber\\
&=& \, 
\lim_{R\to \infty}
\sum_{n_2,\cdots n_{R}=1}^M
\prod_{i=R}^1
\braket{u_{\mathbf{k}_{i+1},n_{i+1}}|u_{\mathbf{k}_{i},n_{i}}}
\nonumber \\
&=& \,  \bra{u_{\mathbf{k}_{1},n_1}} 
\lim_{R\to \infty}
\prod_{i=R}^{2}
\left( \sum_{n_i}^{M} \ket{u_{\mathbf{k}_i,n_i}}\bra{u_{\mathbf{k}_i,n_i}} \right) \ket{u_{\mathbf{k}_{1},n_{1}}},
\end{eqnarray}
where the path $l$ is sampled into $R$ momenta $\mathbf{k}_i$, $i=1,\cdots, R$, and the limit $R\to\infty$ is taken such that the distance between any two neighboring momentum points goes to zero.
Further, $\mathbf{k}_{1}=\mathbf{k}_{R+1}$ are the initial and final momenta along the loop, respectively, on which the Wilson loop matrix depends. 

By the last line of Eq.~\eqref{eq: WLNumericalFormula} it becomes clear that $W[l]$ is gauge \emph{covariant}, that is, transforms as an operator under a general gauge transformation $S(\mathbf{k}) \in U(M)$ of the occupied subspace given by $\ket{u_{\mathbf{k}}} \rightarrow S(\mathbf{k}) \ket{u_{\mathbf{k}}}$, only for a closed loop $l$ (the case where $l$ is non-contractible is also referred to as the Zak phase). However, the Wilson loop \emph{spectrum} for a closed loop is gauge \emph{invariant}, that is, the eigenvalues of $W[l]$ are not affected by gauge transformations (note that they also do not depend on the choice of $\mathbf{k}_i = \mathbf{k}_f$) and may therefore carry physical information. We will show in the following that this is indeed the case: the Wilson loop spectrum is related to the spectrum of the position operator projected into the space of occupied bands.

To proceed, we consider a geometry where $l$ is parallel to the $x$ coordinate axis, and winds once around the BZ. Let $\vv{k}$ denote the $(d-1)$ dimensional vector of remaining good momentum quantum numbers. Then $W(\vv{k})$
is labelled by these remaining momenta. Denote by $\mathrm{exp}(\mathrm{i}\theta_{\alpha,\vv{k}})$, $\alpha=1,\cdots, M$, the eigenvalues of $W(\vv{k})$. The set of phases $\{\theta_{\alpha,\vv{k}}\}$ forms a band structure in the $(d-1)$ dimensional BZ and is often equivalently referred to as the Wilson loop spectrum. Note that all $\theta_{\alpha,\vv{k}}$ are only defined modulo $2\pi$, which makes the Wilson loop spectrum inherently different from the spectrum of a physical Bloch Hamiltonian.

The spectral equivalence we will show relates the eigenvalues of the operator $(-\mathrm{i}/2\pi)\,\mathrm{log} [W(\vv{k})]$ with those of the projected position operator
\begin{equation}
P(\vv{k})\hat{x}P(\vv{k}),
\end{equation}
where the projector $P(\vv{k})$ onto all occupied band eigenstates along $l$ (i.e., all states with wave vector $\vv{k}$) is given by
\begin{equation}
\label{eq: projectorAlongX}
P(\vv{k})=\sum_n^M
\int_{-\pi}^\pi \frac{\mathrm{d} k_x}{2\pi}| \psi_{\mathbf{k},n}\rangle\langle \psi_{\mathbf{k},n}|,
\end{equation}
while the states $\ket{\psi_{\mathbf{k},n}}$ are given by Eq.~\eqref{eq: fullEigenstates}.
The eigenvalues of the projected position operator have the interpretation of the charge centers in the ground state of the Hamiltonian considered, while the eigenstates are known as hybrid Wannier states, which are localized in the $x$-direction and plane waves perpendicular to it~\cite{Souza12}.

To prove the equivalence, we start with the eigenfunctions of $P(\vv{k})\hat{x}P(\vv{k})$, which satisfy 
\begin{equation}
\left[P(\vv{k})\hat{x}P(\vv{k})-\frac{\tilde{\theta}_{\vv{k}}}{2\pi}\right]|\Psi_{\vv{k}}\rangle=0.
\label{eq: eigenvalueEq}
\end{equation}
Note that there are $M$ eigenvectors, the form of the corresponding eigenvalues $\tilde{\theta}_{\alpha,\vv{k}}/(2\pi)$, $\alpha=1,\cdots, M$ has been chosen for later convenience and in particular has not yet been logically connected to the $\theta_{\alpha,\vv{k}}$ making up the Wilson loop spectrum (however, we will do so shortly).
An eigenfunction can be expanded as
\begin{equation}
|\Psi_{\vv{k}}\rangle=\sum_{n}^M\int\mathrm{d}k_x \, f_{\vv{k},n}(k_x)|\psi_{\mathbf{k},n}\rangle,
\end{equation}
where the coefficients $f_{\vv{k},n}$ satisfy the equation
\begin{eqnarray}
&&\langle\psi_{\mathbf{k},n}|P(\vv{k})\hat{x}P(\vv{k})|\Psi_{\vv{k}}\rangle \nonumber\\
&=&
\sum_m \int \mathrm{d} \tilde{k}_x \braket{\psi_{\mathbf{k},n} | (\mathrm{i} \partial_{\tilde{k}_x}) f_{\vv{k},m}(\tilde{k}_x) | \psi_{\mathbf{\tilde{k}},m}} \nonumber\\
&=& \sum_m \int \mathrm{d} \tilde{k}_x \, \mathrm{i} \frac{\partial f_{\vv{k},m}(\tilde{k}_x)}{\partial \tilde{k}_x} (\delta_{m,n} \delta_{\tilde{k}_x, k_x}) \\&&
+ \sum_m \int \mathrm{d} \tilde{k}_x f_{\vv{k}, m} (\tilde{k}_x) \int \frac{\mathrm{d} x}{2\pi} \braket{u_{\mathbf{k},n} | e^{- \mathrm{i} k_x x} (\mathrm{i} \partial_{\tilde{k}_x}) e^{\mathrm{i} \tilde{k}_x x} | u_{\mathbf{\tilde{k}},m} } \nonumber\\
&=& \mathrm{i} \frac{\partial f_{\vv{k},n}(k_x)}{\partial k_x} - f_{\vv{k},n}(k_x) \int \frac{\mathrm{d}x}{2\pi} x  
+ \mathrm{i} \sum_m \int \mathrm{d} \tilde{k}_x f_{\vv{k}, m}(\tilde{k}_x) \int \frac{\mathrm{d} x}{2\pi} e^{-\mathrm{i} (k_x - \tilde{k}_x) x} \braket{u_{\mathbf{k}, n} | \partial_{k_x} | u_{\mathbf{\tilde{k}}, m}} \nonumber \\
&=& \mathrm{i} \frac{\partial f_{\vv{k},n}(k_x)}{\partial k_x} + \mathrm{i} \sum_m f_{\vv{k}, m} (k_x) \braket{u_{\mathbf{k}, n} | \partial_{k_x} | u_{\mathbf{k}, m}}
\nonumber\\
&=&\mathrm{i}\frac{\partial f_{\vv{k},n}(k_x)}{\partial k_x}+\mathrm{i}\sum_m^M A_{x;n,m}(\mathbf{k}) f_{\vv{k},m}(k_x).
\end{eqnarray}
(Note that we have to assume an appropriate regularization to make the term $\int \mathrm{d}x \, x$ vanish in this continuum calculation, reflecting the ambiguity in choosing the origin of the coordinate system.)
Then, integrating the resulting Eq.~\eqref{eq: eigenvalueEq} for $f_{\vv{k},n}(k_x)$, we obtain
\begin{equation}
f_{\vv{k},n}(k_x)
=e^{-\mathrm{i}(k_x-k_x^0)\tilde{\theta}_{\vv{k}}/(2\pi)}
\sum_m^M
\overline{\mathrm{exp}}
\left[
-\int_{k_x^0}^{k_x}  \mathrm{d}\tilde{k}_x A_{x}(\tilde{k}_x,\vv{k})
\right]_{n,m}
f_{\vv{k},m}(k_x^0).
\end{equation}
We now choose $k_x=k_x^0+2\pi$.
Periodicity of $f_{\vv{k},m}(k_x^0)$ as $k_x^0 \rightarrow k_x^0 + 2 \pi$ yields (choosing $k_x^0 = \pi$ without loss of generality)
\begin{equation}
\sum_m^M W(\vv{k})_{n,m} f_{\vv{k},m}(\pi)
=e^{\mathrm{i}\tilde{\theta}_{\vv{k}}} f_{\vv{k},n}(\pi),
\end{equation}
showing that the expansion coefficients of an eigenstate of $P(\vv{k})\hat{x}P(\vv{k})$ with eigenvalue $\tilde{\theta}_{\vv{k}}/(2\pi)$ form eigenvectors of $W(\mathbf{k})$ with eigenvalues $e^{\mathrm{i}\tilde{\theta}_{\vv{k}}}$. This establishes the spectral equivalence $\tilde{\theta}_{\vv{k}} = \theta_{\vv{k}}$.

Note that there are $M$ eigenvalues of the Wilson loop, while the number of eigenvalues of $P(\vv{k})\hat{x}P(\vv{k})$ is extensive in the system size.
Indeed, for each occupied band (i.e., every Wilson loop eigenvalue $\theta_{\alpha,\vv{k}}$, $\alpha=1,\cdots,M$) there exists a ladder of eigenvalues of the projected position operator
\begin{equation}
\frac{\theta_{\alpha,\vv{k},X}}{2\pi}=\frac{\theta_{\alpha,\vv{k}}}{2\pi}+X,\qquad
X\in\mathbb{Z},\qquad \alpha=1,\cdots, M.
\end{equation}
Notice that we have set the lattice spacing in the $x$-direction to 1 for convenience here and in the following.

The eigenstates of the projected position operator are hybrid Wannier states which are maximally localized in $x$ direction, but take on plane wave form in the perpendicular directions. Note that since the eigenvalues of $W(\mathbf{k})$ along any non-contractible loop of $\mathbf{k}$ in the BZ define a map $S^1 \rightarrow U(1) \cong S^1$, their winding number, which is necessarily an integer, can, given additional crystalline symmetries, provide a topological invariant that cannot be changed by smooth deformations of the system's Hamiltonian. To familiarize the reader with the concepts introduced above, we now present the properties of Wilson loop spectra in the context of three simple models.

\subsubsection{Example: Su-Schrieffer-Heeger model}
\label{sec: SSHexample}
\begin{figure}[t]
\begin{center}
\includegraphics[width=0.75 \textwidth,page=1]{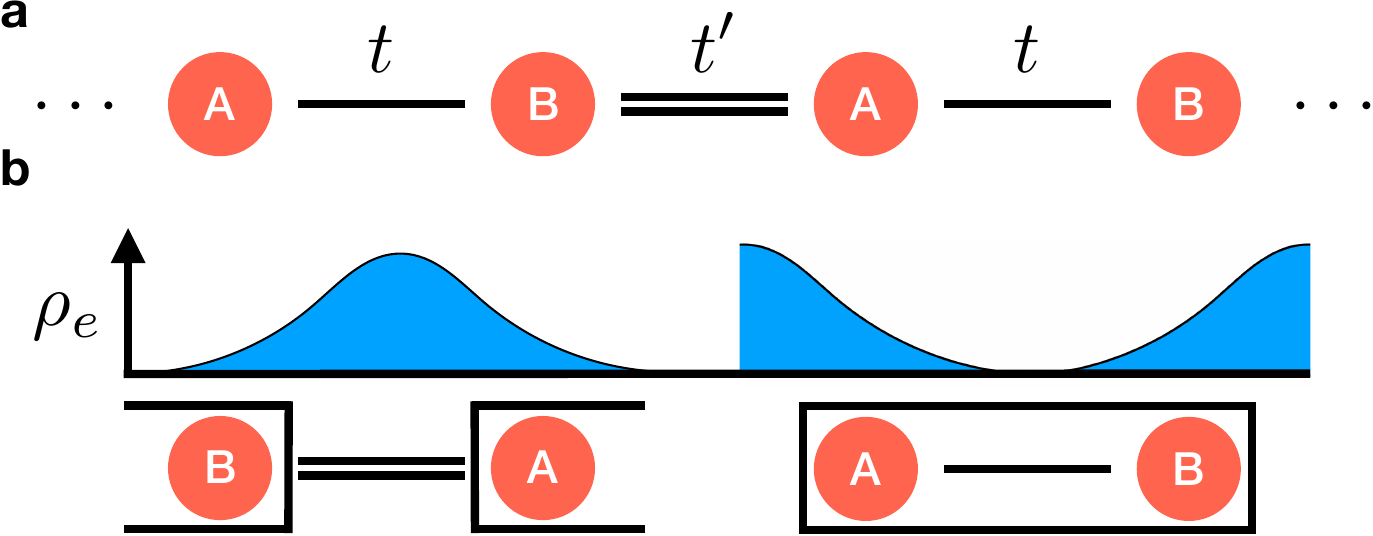}
\caption{
Su-Schrieffer-Heeger model. \textbf{a}  The model consists of electrons hopping on a dimerized chain with alternating hopping strengths $t$ and $t'$. For the case of $t' > t$, the model is in its topological phase, which is adiabatically connected to the special case $t' \neq 0$, $t=0$. In this limit the presence of gapless edge modes is evident when the chain is cut after a full unit cell. \textbf{b}  The polarization is a measure of where charges sit in the unit cell. Shown is the case $\mathsf{P} = 1/2$, where the charge center is displaced by exactly half a lattice spacing. When cutting the system after a full unit cell, the edge hosts a state of charge $1/2$. This is the simplest example of charge fractionalization in topological condensed matter systems.
}
\label{fig: SSH}
\end{center}
\end{figure}

One of the simplest examples of a topological phase is exemplified by the Su-Schrieffer-Heeger (SSH) model, initially devised to model polyacetylene. It describes electrons hopping on a 1D dimerized lattice with two sites $A$ and $B$ in its unit cell (see Fig~\ref{fig: SSH}\textbf{a}). In momentum space, the Bloch Hamiltonian reads
\begin{equation}
\label{eq: SSHdef}
\mathcal{H}(k) = \begin{pmatrix}
0&t+t' e^{\mathrm{i} k}\\
t+t' e^{-\mathrm{i} k}&0
\end{pmatrix}.
\end{equation}
The model has an inversion symmetry $I \mathcal{H}(k) I^{-1} = \mathcal{H}(-k)$, with $I=\sigma_x$. Since it does not couple sites $A$ to $A$ or $B$ to $B$ individually, it furthermore enjoys a chiral or sublattice symmetry $C \mathcal{H}(k) C^{-1} = -\mathcal{H}(k)$ with $C = \sigma_z$. [Notice some abuse of language here: The chiral symmetry is not a ``symmetry'' in the sense of a commuting operator on the level of the first quantized Bloch Hamiltonian. Still, as a mathematical fact, this chiral symmetry can be helpful to infer and protect the existence of topological boundary modes.] While a standard discussion of the SSH model would focus on the chiral symmetry and its role in protecting topological phases, here we will first consider the implications of the crystalline inversion symmetry. It will be useful to note that the spectrum is given by $E = \pm \sqrt{t^2 + t'^2 + 2 t t' \cos{k}}$ with a gap closing at $k = \pi$ for $t = t'$ and at $k=0$ for $t=-t'$.

Let us start by calculating the Wilson loop for the case where $(t,t')=(0,1)$. The eigenvectors of $\mathcal{H}(k)$ are then given by
\begin{equation}
\ket{u_{k,1}} = \frac{1}{\sqrt{2}} \begin{pmatrix} -e^{\mathrm{i} k} \\ 1 \end{pmatrix}, \quad \ket{u_{k,2}} = \frac{1}{\sqrt{2}} \begin{pmatrix} e^{\mathrm{i} k} \\ 1 \end{pmatrix},
\end{equation}
with energies $-1$ and $+1$, respectively. Since the occupied subspace is one-dimensional in this case, the Berry connection $A(k) = \braket{u_{k,1}| \partial_k | u_{k,1}} = \mathrm{i}/2$ is Abelian and given by just a purely imaginary number (remember that it is anti-Hermitian in general).
We thus obtain
\begin{equation}
\mathsf{P} := -\frac{\mathrm{i}}{2 \pi} \log W = -\frac{\mathrm{i}}{2 \pi} \int_0^{2 \pi} A(k) \mathrm{d}k = \frac{1}{2}.
\end{equation}
The physical interpretation of $\mathsf{P}$ is given within the modern theory of polarization (see Ref.~\cite{SpaldinLectures} for a pedagogical introduction) as that of a bulk electrical dipole moment or charge polarization, which is naturally only defined modulo $1$ since the coordinate of a center of charge on the lattice is only defined up to a lattice translation (remember that we have chosen the lattice spacing $a=1$). It is directly connected to the Wilson loop spectrum $\theta_{\alpha,\vv{k}}$ by a rescaling which makes sure that the periodicity of the charge centers defined in this way is that of the real-space lattice. See also Fig.~\ref{fig: SSH}\textbf{b}.

The result $\mathsf{P}=1/2$ is by no means accidental: In fact, since the inversion symmetry reverses the path of integration in $W$, but leaves inner products such as $A(k)$ invariant, the Wilson loop eigenvalues of an inversion symmetric system satisfy $e^{\mathrm{i}\theta} = e^{-\mathrm{i}\theta}$ (see also Sec.~\ref{sec: symm on Wilson} below). This requires that $\mathsf{P}$ be quantized to $0$ ($\theta=0$) or $1/2$ ($\theta=\pi$) in the Abelian case. This is a first example where a crystalline symmetry such as inversion, which acts non-locally in space, protects a topological phase by enforcing the quantization of a topological invariant to values that cannot be mapped into one another by an adiabatic evolution of the corresponding Hamiltonian.
Note that since the eigenstates for the parameter choice $(t,t')=(1,0)$ do not depend on $k$, we immediately obtain $\mathsf{P} = 0$ for this topologically trivial case.

By these considerations it is clear that in fact the full parameter regime where $t<t'$ is topological, while the regime $t>t'$ is trivial. This is because it is possible to perform an adiabatic interpolation from the specific parameter choices $(t,t')\in\{(0,1),(1,0)\}$ considered above to all other values as long as there is no gap closing and no breaking of inversion symmetry, which is true provided that the line $t=t'$ is avoided in parameter space.

In general, a topological phase comes with topologically protected gapless boundary modes on boundaries which preserve the protecting symmetry. For inversion symmetry, however, there are no boundaries satisfying this requirement. Even though the model at $(t,t')=(0,1)$ has zero-mode end states [since in this case, $\mathcal{H}(k)$ does not act at all on the $A$ ($B$) site in the unit cell at the left (right) edge of the sample], these modes can be removed from zero energy by generic local perturbations even without a bulk gap closing. To protect the end modes, we need to invoke the chiral symmetry, which implies that an eigenstate at any energy $E$ is paired up with an eigenstate at energy $-E$. Eigenstates of the chiral symmetry can then only appear at $E=0$. A spatially and spectrally isolated boundary mode at $E=0$ can thus not be removed by perturbations that retain the chiral symmetry. In conclusion, topological crystalline phases in 1D have no protected boundary degrees of freedom as long as we do not include further local symmetries.

In fact, in the presence of chiral symmetry, the above discussion can be generalized to arbitrary 1D models. In the eigenbasis of $C$, we can write any Hamiltonian with chiral symmetry in the form
\begin{equation}
\mathcal{H}(k) = \begin{pmatrix}
0&q(k)\\
q^\dagger(k)&0
\end{pmatrix},
\end{equation}
where for the SSH model the matrix $q(k)$ was given by just a complex number, and in general we choose it to be a unitary matrix by an adiabatic deformation of the Hamiltonian.
The chiral symmetry allows for the definition of a winding number
\begin{equation}
\label{eq: chiralWindingNumber}
\nu=\frac{\mathrm{i}}{2\pi}
\int\mathrm{d}k \, \mathrm{Tr} \left[q(k)\partial_kq^\dagger(k)\right]
\in \mathbb{Z}.
\end{equation}
This winding number is one of the topological invariants alluded to in Sec.~\ref{sec: motivation} and is only valid when chiral symmetry is present. We can make contact with the overarching concept of Wilson loops by calculating the connection
\begin{equation}
A=\frac{1}{2} q(k)\partial_kq^\dagger(k).
\end{equation}
Thus the Wilson loop eigenvalues $e^{\mathrm{i} \theta_\alpha}$ satisfy
\begin{equation}
\label{eq: wilsonChiralRelation}
\frac{1}{2\pi}\sum_\alpha \theta_\alpha=\frac{\nu}{2}\ \mathrm{mod}\ 1.
\end{equation}
In particular, in the Abelian case, chiral symmetry thus implies the quantization of $\mathsf{P}$ to half-integer values, just as inversion symmetry did it above. An important distinction to be made is that with inversion symmetry, we have a $\mathbb{Z}_2$ topological classification ($\mathsf{P}$ can be either $0$ or $1/2$), while with chiral symmetry the winding number allows for a $\mathbb{Z}$ classification.

\begin{figure}[t]
\begin{center}
\includegraphics[width=0.75 \textwidth,page=2]{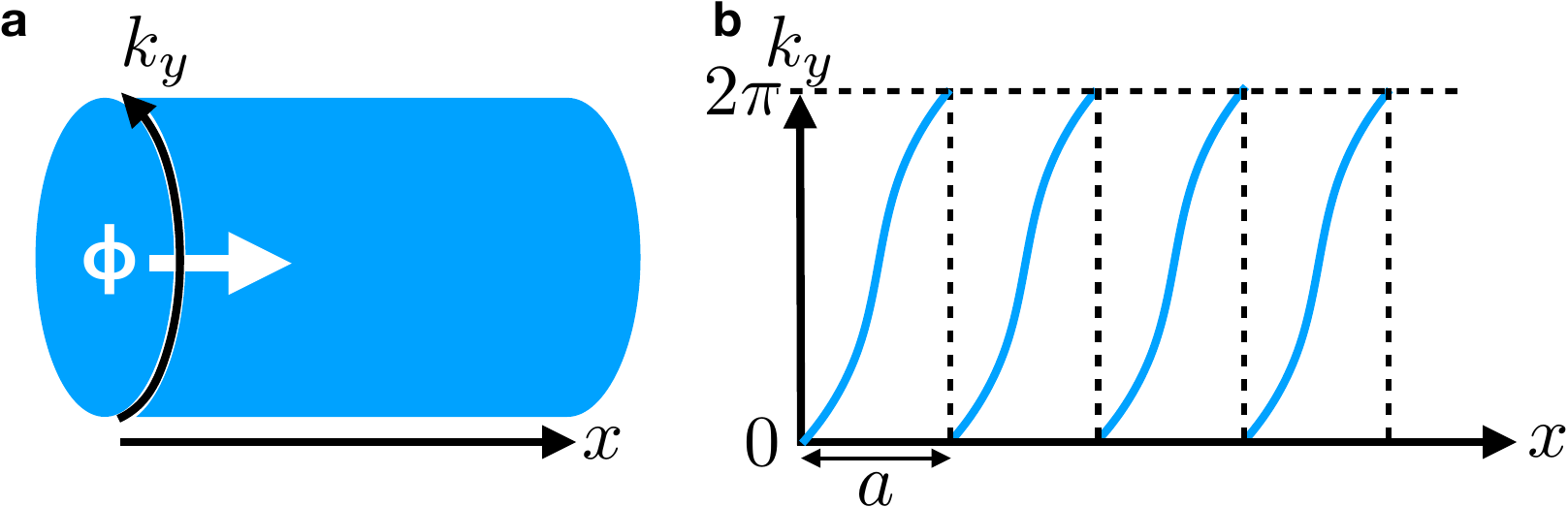}
\caption{
Chern insulator geometry and charge center flow. \textbf{a}  To calculate the Hall conductivity, we consider a gedanken experiment where the $y$-direction of 2D space is compactified while we retain open boundary conditions in the $x$-direction. In particular, the translational symmetry along $y$ allows for the introduction of the momentum $k_y$ as a good quantum number to label blocks of the Hamiltonian and eigenstates. The Hall conductivity is then equal to the amount of charge transported along the $y$-direction in a single adiabatic cycle of flux insertion, where the inserted flux $\phi$ varies over time from $0$ to $2\pi$. \textbf{b}  Charge center flow corresponding to a Chern number $C = 1$. Since one cycle of flux insertion corresponds to tuning $k_y$ from 0 to $2 \pi$, we see that in one such cycle the charge center crosses exactly one unit cell.
}
\label{fig: ChernInsulator}
\end{center}
\end{figure}

\subsubsection{Example: Chern insulator}
Another paradigmatic example of a topologically protected phase is given by the (integer) quantum Hall effect of electrons subject to a perpendicular magnetic field in 2D continuous space. Here, we study its lattice realization, also called the quantum anomalous Hall effect or Chern insulator. We consider a 2D square lattice with open boundary conditions in $x$-direction and periodic boundary conditions in $y$-direction, retaining the momentum $k_y$ as good quantum number.

To find an expression for the Hall conductivity for any Hamiltonian we could put on this lattice in terms of Wilson loops, let us perform a thought experiment where we roll up the $y$ direction to form a cylinder of circumference $L$ (see Fig.~\ref{fig: ChernInsulator}\textbf{a}). Threading a magnetic flux $\phi$ along the $x$ direction through this cylinder, amounts to the replacement $k_y\to k_y+\phi$ by a Peierls substitution. Note that in our units $\phi = 2\pi$ denotes a single flux quantum.

We now consider a Wilson loop along $x$ direction, labelled by $k_y$ with eigenvalues $e^{\mathrm{i}\theta_{\alpha,k_y}}$. The derivative $\partial_{k_y}\theta_{\alpha,k_y}$ of the $\alpha$-th Wilson loop eigenvalue is by the interpretation in terms of the modern theory of polarization explained in the previous section simply the `velocity' in $x$-direction of the $\alpha$-th charge center at `time' $k_y$. Integrating over $k_y$, i.e., adiabatically performing a flux insertion from $\phi = 0$ to $\phi = 2\pi$ (which brings the system back to its initial state), gives the full Hall conductivity as $2 \pi$ (or, if $e$ and $\hslash$ are reinstated, $e^2/h$) times
\begin{equation}
\label{eq: Laughlin}
C = \sum_\alpha^M\int\frac{\mathrm{d}k_y}{2\pi}\, \partial_{k_y}\theta_{\alpha,k_y},
\end{equation}
where $C$ is known as the Chern number. To see how this formula works, note that the Hall conductivity is equal to the amount of charge transported in $y$ direction under the adiabatic insertion of a single flux quantum. Since we can only transport an integer number of charge around the cylinder in one such evolution (at least in the non-interacting systems we are considering here), $C$ is necessarily quantized.

Making use of the relation $\sum_\alpha^M \theta_{\alpha,k_y} = \mathrm{i}\int \mathrm{d} k_x \mathrm{Tr} A_x(k_x,k_y)$, which follows from Eq.~\eqref{eq: WilsonLoop}, and requiring $C$ to be gauge invariant, we can generalize Eq.~\eqref{eq: Laughlin} to
\begin{equation}
\label{eq: ChernWilsonDef}
C= -\frac{\mathrm{i}}{2\pi}
\int\mathrm{d}^2{\mathbf{k}}
\left[\partial_{k_x}\mathrm{Tr} A_{y}(\mathbf{k})
-\partial_{k_y}\mathrm{Tr} A_{x}(\mathbf{k})\right].
\end{equation}
The equality is directly seen in a gauge in which the integral of the first term $\partial_{k_x}\mathrm{Tr} A_{y}(\mathbf{k})$ does not contribute, which we have implicitly been working in (note that $\mathbf{A}$ here denotes the Berry connection, \emph{not} the electromagnetic gauge field).
The Chern number is thus the net number of charge centers crossing a given $x$ position in the full $k_y$ BZ. In the Wilson loop picture, it just corresponds to the winding number of the $x$-direction Wilson loop eigenvalues as $k_y$ is varied along a non-contractible loop in the BZ, which is of course quantized (see Fig.~\ref{fig: ChernInsulator}\textbf{b}). While the Chern number is normally defined by employing the concept of Berry curvature, we have shown here that it may be equivalently expressed in terms of the spectral flow of Wilson loop eigenvalues as described at the end of Sec.~\ref{sec: WilsonLoopPositionOperator}.

\subsubsection{Example: Time-reversal invariant topological insulator}
\label{sec: symm on Wilson}
Here, we explore the constraints imposed by time-reversal or inversion symmetries on Wilson loops. These symmetries protect topological insulators in two and three dimensions.
In the presence of an anti-unitary time-reversal symmetry $\Theta$, a Wilson loop $W_{2\pi\leftarrow0}(\mathbf{k})$ along the $x$-direction, with $k_x$ running from $0$ to $2\pi$, transforms as
\begin{equation}
\begin{split}
\Theta W_{2\pi\leftarrow0}(\mathbf{k}) \Theta^{-1}
=&W^*_{0\leftarrow2\pi}(-\mathbf{k})\\
=&W^{\mathsf{T}}_{2\pi\leftarrow 0}(-\mathbf{k})\\
\Rightarrow& \qquad \theta_\alpha(\mathbf{k})=\theta_\alpha(-\mathbf{k}).
\end{split}
\end{equation}
In particular, in a spinful system where $\Theta^2=-1$, the representation of the time-reversal operation on the Wilson loop retains its property to square to $-1$, so that there is a Kramers degeneracy not only in the energy spectrum, but also in the Wilson loop spectrum. We thus recover the $\mathbb{Z}_2$ classification of 
2D time-reversal invariant topological insulators from the spectral flow in the Wilson loop eigenvalues: Either the bands emerging from individual Kramers pairs connect back to the same pairs as $\mathbf{k}$ evolves along a non-contractible loop in the BZ, or they split up to connect to separate pairs.

Inversion $I$ generates the following spectral pairing
\begin{equation}
\begin{split}
I W_{2\pi\leftarrow0}(\mathbf{k}) I^{-1}
=&W_{0\leftarrow2\pi}(-\mathbf{k})\\
=&W^{\dagger}_{2\pi\leftarrow 0}(-\mathbf{k})\\
\Rightarrow& \qquad \theta_\alpha(\mathbf{k})=-\theta_\alpha(-\mathbf{k}).
\end{split}
\end{equation}

The combination of inversion $I$ and time-reversal $\Theta$ then leads to a `chiral symmetry' for the Wilson loop
\begin{equation}
\begin{split}
I\Theta W_{2\pi\leftarrow0}(\mathbf{k})\Theta^{-1} I^{-1}
=&W^*_{2\pi\leftarrow0}(\mathbf{k})\\
\Rightarrow&\qquad \theta_\alpha(\mathbf{k})=-\theta_\alpha(\mathbf{k}).
\end{split}
\end{equation}

Note that as the $\theta_\alpha(\mathbf{k})$ are only defined modulo $2\pi$, we can have unidirectional flow in the Wilson loop spectrum: in the simplest case, in 2D we could have a single Wilson loop band which winds once along the $\theta$-direction as $k_y$ goes from $0$ to $2 \pi$. This is in stark contrast to energy spectra, in which every unidirectionally dispersing band is paired up with a band going into the opposite direction so that the net chirality of the spectrum is always zero, a result which follows from the Nielsen-Ninomiya theorem under physically realistic circumstances such as locality~\cite{Nielsen1983}.

\subsection{Bulk-boundary correspondence}
As alluded to in Sec.~\ref{sec: motivation}, Wilson loops not only provide a convenient formulation of many topological invariants, but are also in one-to-one correspondence with the boundary degrees of freedom of the system considered. We will now show that indeed the spectrum of a Hamiltonian in the presence of a boundary is smoothly connected to the spectrum of its Wilson loop along the direction perpendicular to the boundary. Note that since the Wilson loop is determined entirely by the bulk Bloch Hamiltonian, this relation provides an explicit realization of the bulk-boundary correspondence underlying all topological phases~\cite{FidkowskiWilsonBulkBoundary11}.

\begin{figure}[t]
\begin{center}
\includegraphics[width=0.75 \textwidth,page=4]{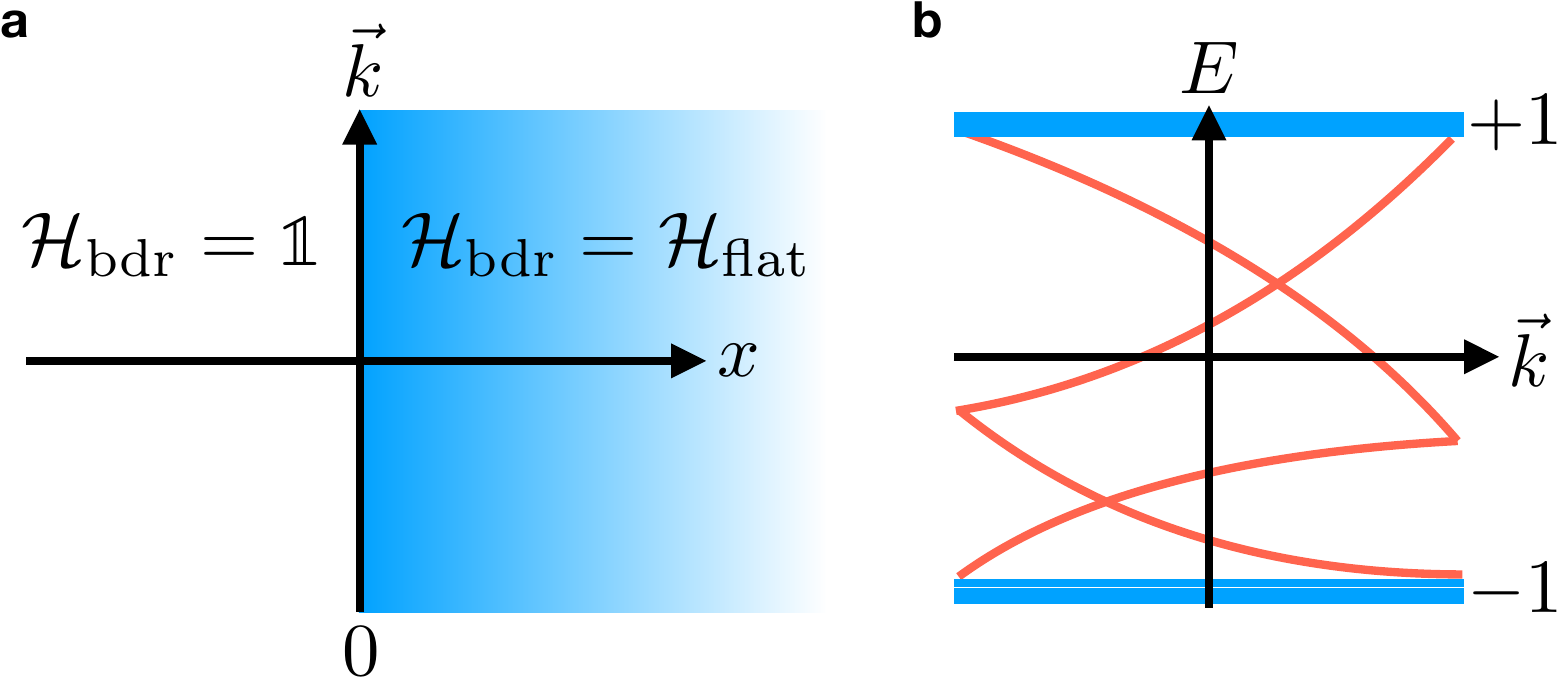}
\caption{
Real space setup and generic spectrum of $\mathcal{H}_{\mathrm{bdr}}$. \textbf{a}  For $V_0(x)$ as given by Eq.~\eqref{eq: V0}, $\mathcal{H}_{\mathrm{bdr}}$ varies discontinuously from a trivial projector in the domain $x<0$ to $\mathcal{H}_\mathrm{flat}$ in the domain $x>0$. Translational symmetry along $x$ is thus broken, however it is preserved along all perpendicular directions, which still have good momentum quantum numbers $\mathbf{k}$. \textbf{b}  The spectrum of $\mathcal{H}_{\mathrm{bdr}}$ has accumulation points at $\pm 1$, stemming from the semi-infinite regions to the left and right of the domain wall, and a discrete set of bands in between, coming from the finite domain wall region.}
\label{fig: BBInterpolation}
\end{center}
\end{figure}

We consider a semi-infinite slab geometry with a single edge of the system at $x = 0$, while keeping $\mathbf{k}$ as good quantum numbers. 
From a topological viewpoint, the actual energetics of the band structure are irrelevant, and we can always deform the Hamiltonian for the sake of clarity to a spectrally flattened Hamiltonian where all bands above and below the gap are at energy $+1$ and $-1$, respectively, without closing the gap. It is therefore enough to work with
\begin{equation}
\mathcal{H}_{\mathrm{flat}}(\mathbf{k})=1-2 P(\mathbf{k})
\end{equation}
to model the bulk system. Here, $P(\mathbf{k})$ as defined in Eq.~\eqref{eq: projectorAlongX}, repeated here for convenience,
\begin{equation}
P(\mathbf{k})=\sum_n^M
\int_{-\pi}^\pi \frac{\mathrm{d} k_x}{2\pi}| \psi_{\mathbf{k},n}\rangle\langle \psi_{\mathbf{k},n}|,
\end{equation}
is the projector onto the occupied subspace for a given $\mathbf{k}$. Note that $\mathcal{H}_{\mathrm{flat}}(\mathbf{k})$ actually has the same eigenvectors as the original Hamiltonian.
To model a system with boundary, we use
\begin{equation}
\mathcal{H}_{\mathrm{bdr}}(\mathbf{k})=P(\mathbf{k})V_0(\hat{x})P(\mathbf{k})+1- P(\mathbf{k}),
\end{equation}
with 
\begin{equation}
\label{eq: V0}
V_0(x)=
\begin{cases}
1&x<0\\
-1&x>0
\end{cases}
\end{equation}
so that we have $\mathcal{H}_{\mathrm{bdr}}(\mathbf{k})\to \mathcal{H}_{\mathrm{flat}}(\mathbf{k})$ for $x\to+\infty$ and $\mathcal{H}_{\mathrm{bdr}}(\mathbf{k})\to 1$ for $x\to-\infty$ (see Fig.~\ref{fig: BBInterpolation}\textbf{a}).
The latter limit corresponds to a description of the vacuum with the chemical potential chosen so that no electron states will be occupied, which we take to be the topologically trivial limit.

Since we take space to be infinitely extended away from the domain wall at $x=0$, the spectrum of $\mathcal{H}_{\mathrm{bdr}}(\mathbf{k})$ includes the spectrum of $\mathcal{H}_{\mathrm{flat}}(\mathbf{k})$, given by $\pm 1$ since $P^2(\mathbf{k}) = P(\mathbf{k})$, as well as that of the operator $1$, trivially given by $+1$. The boundary region is of finite extent and can therefore contribute only a finite number of midgap states as the system has exponentially decaying correlations on either side of the boundary. There are therefore spectral accumulation points at $\pm1$, but otherwise we are left with a discrete spectrum (see Fig.~\ref{fig: BBInterpolation}\textbf{b}). We will focus on this part of the spectrum.

We will now deform the spectrum of $\mathcal{H}_{\mathrm{bdr}}(\mathbf{k})$ to that of $(-\mathrm{i}/2\pi)\,\mathrm{log} [W(\mathbf{k})]$ by considering an evolution that takes $P(\mathbf{k})V_0(\hat{x})P(\mathbf{k})$ to $P(\vv{k})\hat{x}P(\vv{k})$, the eigenvalues of which were previously shown to be directly related to those of $(-\mathrm{i}/2\pi)\,\mathrm{log} [W(\mathbf{k})]$.
The deformation is continuous in $\mathbf{k}$ and therefore preserves both discreteness of the spectrum as well as its topological properties. An example for this interpolation is given by
\begin{equation}
V_t(x)
=
\begin{cases}
-\frac{x}{t}&\mathrm{for}\ |x|<t/(1-t)\\
-\frac{\mathrm{sgn} (x)}{1-t}&\mathrm{for}\ |x|\geq t/(1-t)
\end{cases},
\qquad 0\leq t\leq1.
\end{equation}
Importantly, for any $t<1$, $P(\mathbf{k})V_t(\hat{x})P(\mathbf{k})$ is a finite rank (finite support) perturbation of $(1-t)^{-1}P(\mathbf{k})V_0(\hat{x})P(\mathbf{k})$, so it will retain the property that the spectrum is discrete. However, the point $t=1$ deserves closer inspection, as $P(\vv{k})\hat{x}P(\vv{k})$ is not a bounded operator. However, we can handle this subtlety by defining 
\begin{equation}
h(r)=
\begin{cases}
r&\mathrm{for} -w<r<w\\
\mathrm{sgn}(r)w &\mathrm{else} 
\end{cases}
\end{equation} 
and considering $h[P(\mathbf{k})V_t(\hat{x})P(\mathbf{k})]$ for some large $w$. The spectrum evolves uniformly continuously from $h[P(\mathbf{k})V_0(\hat{x})P(\mathbf{k})]$ to $h[P(\mathbf{k})V_1(\hat{x})P(\mathbf{k})]$ for any finite $w$~\cite{FidkowskiWilsonBulkBoundary11}.

The topology of the Wilson loop spectrum and the physical boundary spectrum is thus identical. Protected spectral flow in the former implies gapless boundary modes in the latter, as long as the form of the boundary [i.e., $V(x)$] does not break a symmetry that protects the bulk spectral flow.

\subsubsection{Example: Chern insulator spectral flow}
\label{sec: ChernSpecFlowExample}
\begin{figure}[t]
\begin{center}
\includegraphics[width=0.75 \textwidth,page=3]{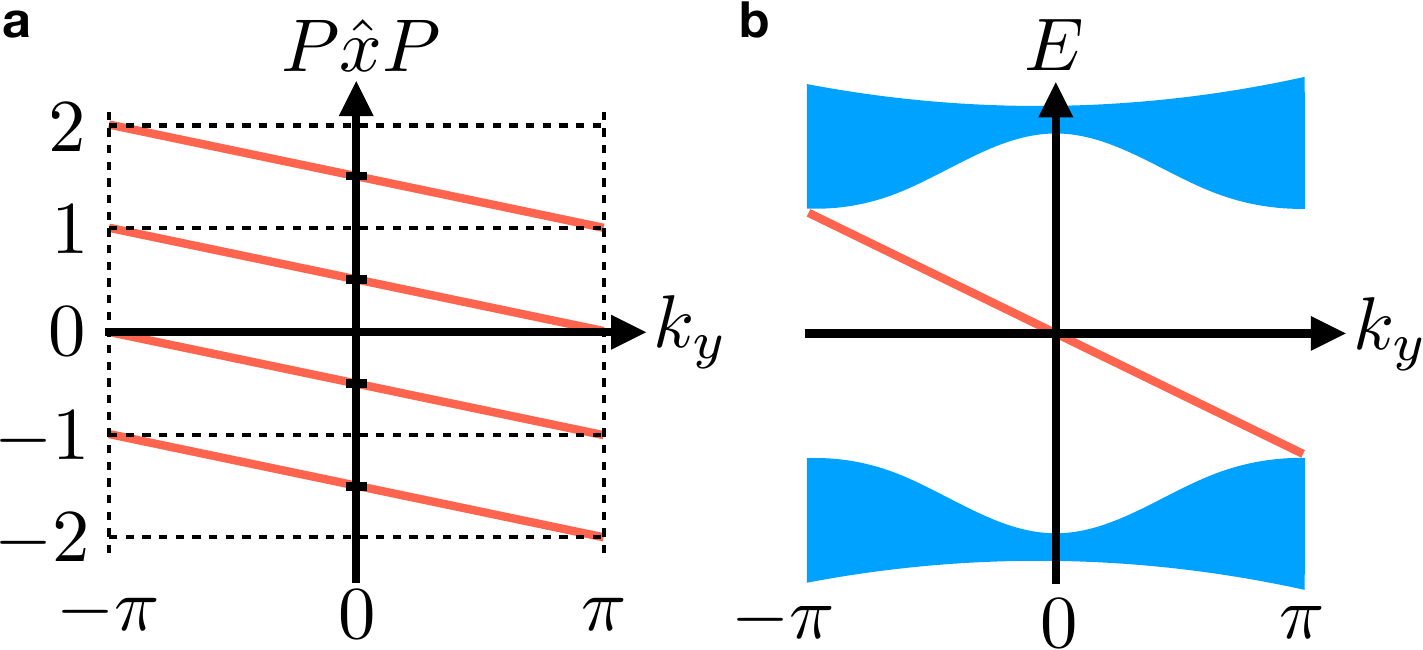}
\caption{
Projected position operator spectral flow and its implications for the boundary Hamiltonian of a Chern insulator. \textbf{a}  The model in Eq.~\eqref{eq: ChernFlow} has two bands, therefore the Wilson loop is Abelian. Due to $C = 1$, by the spectral equivalence derived in Sec.~\ref{sec: WilsonLoopPositionOperator}, the eigenvalues of the projected position operator $P\hat{x}P$ flow from an integer $n$ back to $n-1 = n \mod 1$ exactly once as $k_y$ is varied across a non-contractible loop in the BZ. \textbf{b}  By the bulk-boundary correspondence derived in the present chapter, this implies a single chiral mode crossing the gapped bulk bands when the system is considered in the presence of a boundary termination.
}
\label{fig: ChernSpectralFlow}
\end{center}
\end{figure}

We can obtain a simple Hamiltonian for a Chern insulator in 2D from that of the SSH model in 1D by tuning the latter from its topological to its trivial phase along a perpendicular direction $k_y$ in the BZ. Along the way, we have to make sure that the whole system stays gapped. One Hamiltonian that does the job is given by [compare to Eq.~\eqref{eq: SSHdef}]
\begin{equation}
\label{eq: ChernFlow}
\mathcal{H}(\mathbf{k}) = \begin{pmatrix} \sin k_y & (1-\cos k_y) + e^{\mathrm{i} k_x} \\ (1-\cos k_y) + e^{-\mathrm{i} k_x} & -\sin k_y \end{pmatrix}.
\end{equation}
Here, the part proportional to $\sin k_y$ is the term we added to keep the system gapped at all points in the new 2D BZ.
The Wilson loop we considered in Sec.~\ref{sec: SSHexample}, and with it the polarization $\mathsf{P}$, now becomes a function of $k_y$. We know that $\mathcal{H}(k_x,0)$ corresponds to a topological SSH chain, while $\mathcal{H}(k_x,\pi)$ corresponds to a trivial one, implying $\mathsf{P}(0) = 1/2$ and $\mathsf{P}(\pi) = 0$. Remembering that $\mathsf{P}$ is only defined up to an integer, there are two possibilities for the Wilson loop spectral flow as $k_y$ is varied from $-\pi$ to $+\pi$: Either the Wilson loop bands connect back to themselves trivially after the cycle has come to a close, or they do so only modulo an integer given by the Chern number $C$ in Eq.~\eqref{eq: Laughlin}. To infer which case applies to the model at hand, we can use the relation $\mathcal{H}(k_x,k_y)=-\sigma_3\mathcal{H}(k_x,-k_y)\sigma_3$, which is a combination of chiral symmetry and $y$-mirror symmetry and must also hold in presence of a boundary (if the boundary potential is chosen such that it does not break this symmetry). It dictates that boundary spectra consist of \emph{chiral} modes that connect the SSH spectra at $k_y=0$ and $k_y=\pi$ as shown in Fig.~\ref{fig: ChernSpectralFlow}.
(Note that invoking the combination of chiral and mirror symmetry is only a convenient way to infer the boundary mode connectivity. No symmetry is needed to protect chiral boundary modes.)
This is consistent with the Chern number, which for the model at hand evaluates to $C = 1$. 

\section{Topological crystalline insulators}
Topological crystalline insulators~\cite{Fu11} are protected by spatial symmetry transformations which act non-locally such as mirror or rotational symmetries. They are usually identified with two notions: i) their bulk ground state is not adiabatically connected to an atomic limit without breaking the protecting symmetry. ii) they have gapless boundary modes which can only be gapped out by breaking the respective symmetry.

In fact, properties i) and ii) are not equivalent. We have already seen for the case of the SSH model protected by inversion symmetry that it is possible to have a model featuring i) but not ii). The reason was that although the model in its topological phase (as detectable by, e.g., the Wilson loop) is not adiabatically connected to any atomic limit, there is no boundary which is left invariant by inversion symmetry, and thus no protected edge modes (as long as we do not consider chiral symmetry, which is local and therefore non-crystalline). This is a general feature of topological crystalline insulators: gapless symmetry-protected boundary modes also require the boundary on which they are localized to preserve the corresponding symmetry. In the following, we discuss several pedagogical examples of topological crystalline phases and their invariants.

\subsection{2D topological crystalline insulator}
\label{sec: 2DTCI}
Here we show how crystalline symmetries can enrich the topological classification of band structures. We begin with a model with chiral symmetry in 2D. A natural non-local symmetry in 2D we can add is a mirror symmetry, which leaves an edge invariant. While all 2D systems with just chiral symmetry (class AIII in the tenfold way) are topologically trivial, it will turn out that with mirror symmetry this is no longer the case when we require that mirror and chiral symmetry transformations commute. Note that in contrast, the Chern insulator model we considered in Sec.~\ref{sec: ChernSpecFlowExample} breaks the chiral symmetry of the SSH models from which it was constructed by the gapping term proportional to $\sin k_y$. It therefore belongs to symmetry class A (no symmetries) and can be topological without crystalline symmetries.

The model we consider here is defined by the Bloch Hamiltonian
\begin{eqnarray}
\label{eq: 2DTCISSHModel}
\mathcal{H}(\mathbf{k}) &=& \begin{pmatrix}
0&q(\mathbf{k})\nonumber\\
q^\dagger(\mathbf{k})&0
\end{pmatrix},
\nonumber\\
q(\mathbf{k}) &=& \begin{pmatrix} (1-\cos k_y) + e^{\mathrm{i} k_x} +\lambda & \sin k_y \\ -\sin k_y & (1-\cos k_y) + e^{-\mathrm{i} k_x} -\lambda \end{pmatrix}.
\end{eqnarray}
The symmetry representations are
\begin{eqnarray}
C \mathcal{H}(\mathbf{k}) C^{-1} &=& - \mathcal{H}(\mathbf{k}), \quad M_y \mathcal{H}(k_x, k_y) M_y^{-1} = \mathcal{H}(k_x, -k_y),
\nonumber\\
C &=& \begin{pmatrix} \mathbb{1}_{2 \times 2} & 0 \\ 0 & -\mathbb{1}_{2 \times 2} \end{pmatrix}, \quad M_y = \begin{pmatrix} \sigma_z & 0 \\ 0 & \sigma_z \end{pmatrix},
\end{eqnarray}
where $\mathbb{1}_{2 \times 2}$ denotes the $2 \times 2$ identity matrix and $\lambda$ represents a numerically small perturbation that breaks $M_x$ symmetry. When we calculate the winding number as defined in Eq.~\eqref{eq: chiralWindingNumber} along the path $k_x = 0 \rightarrow k_x = 2\pi$, $k_y = \mathrm{const.}$ in the BZ, we find $\nu(k_y) = 0$ $\forall k_y$. We can most easily see this by evaluating $\nu(0) = 0$ and noting that as the spectrum is gapped throughout the BZ, and the model has chiral symmetry, the result holds for all $k_y$. 

\begin{figure}[t]
\begin{center}
\includegraphics[width=0.75 \textwidth,page=5]{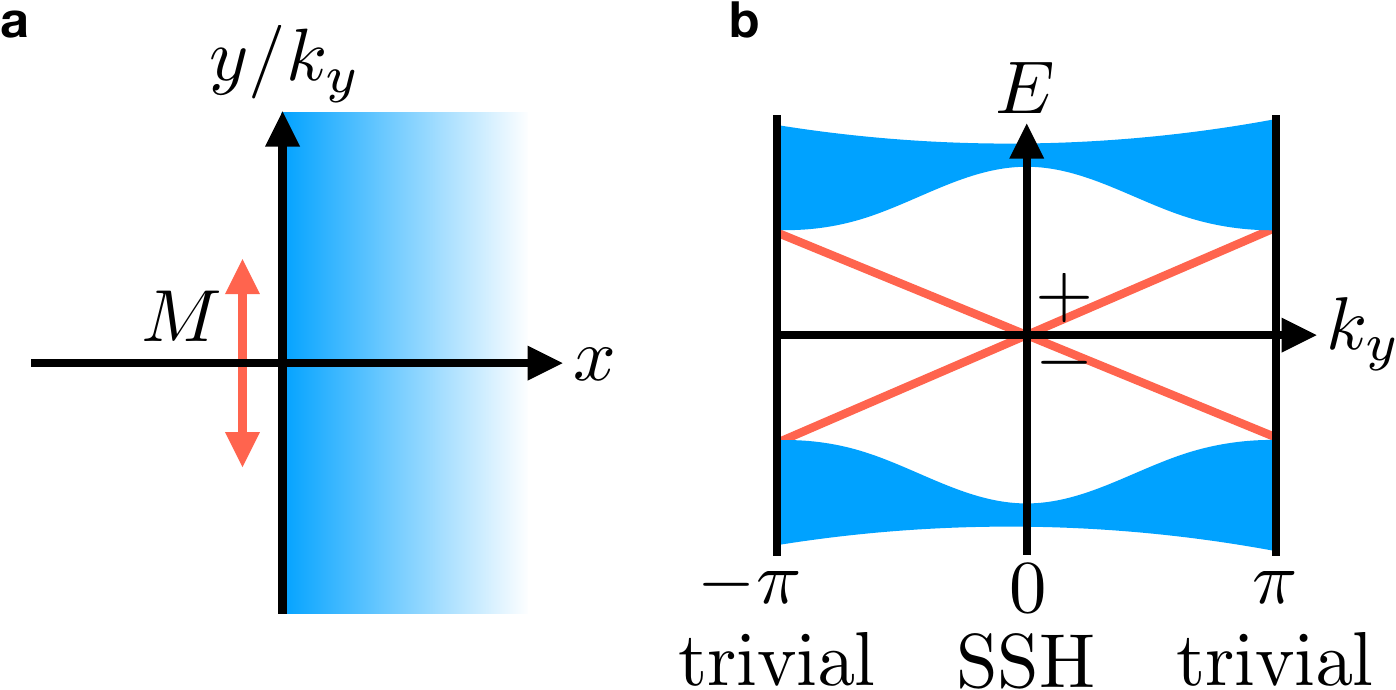}
\caption{
Real space geometry and spectrum of the mirror symmetric 2D model of a chiral symmetric topological crystalline insulator. \textbf{a}  We consider a geometry where the system is terminated in $x$-direction but periodic in $y$-direction, retaining $k_y$ as momentum quantum number. In particular, note that the surface in this semi-infinite slab geometry is mapped onto itself by the $M = M_y$ mirror symmetry, and therefore hosts gapless modes stemming from the nontrivial topology of the bulk. \textbf{b}  Schematic spectrum of the model given by Eq.~\eqref{eq: 2DTCISSHModel} in the presence of the bulk termination in $x$-direction. There are two counter-propagating chiral modes which are necessarily crossing at $k_y = 0$ due to the chiral symmetry. At that point, they are also eigenstates of the mirror symmetry with eigenvalue $\pm 1$, respectively. They are therefore protected from hybridization by the mirror symmetry.
}
\label{fig: MirrorSSH2D}
\end{center}
\end{figure}

In the presence of mirror symmetry, however, we can refine the topological characterization. Since in our case mirror symmetry satisfies $M_y^2 = 1$, its representation has eigenvalues $\pm 1$. Given any line $l_{M_y}$ in the BZ which is left invariant under the action of $M_y$, the eigenstates $\ket{u_{\mathbf{k},n}}$ of $\mathcal{H}$ on $l_{M_y}$ can be decomposed into two groups, $\{\ket{u_{\mathbf{k},l}^+}\}$ and $\{\ket{u_{\mathbf{k},l'}^-}\}$, with mirror eigenvalue $\pm 1$, respectively. We can define the Wilson loop in each mirror subspace as
\begin{equation} 
W^\pm[l_{M_y}]=\overline{\mathrm{exp}}\left[-\int_{l_{M_y}} \mathrm{d}\mathbf{l}_{M_y}\cdot \mathbf{A}^\pm(\mathbf{k})\right],
\end{equation}
where we have used the mirror-graded Berry connection
\begin{equation}
\mathbf{A}^\pm_{m,n}(\mathbf{k})=\langle u^\pm_{\mathbf{k},m}|\mathbf{\nabla}_{\mathbf{k}}|u^\pm_{\mathbf{k},n}\rangle,\qquad n,m=1,\cdots,M.
\end{equation}
For the two mirror invariant paths $l_{M_y}: k_x = 0 \rightarrow k_x = 2\pi, k_y = 0,\pi$, the mirror-graded topological polarization invariants evaluate to 
\begin{equation}
\mathsf{P}_{M_y}(k_y) = \frac{1}{2} \left[\left(-\frac{\mathrm{i}}{2 \pi} \log W^+(k_y) \right) - \left(-\frac{\mathrm{i}}{2 \pi} \log W^-(k_y) \right) \right] = \begin{cases} 1/2 & k_y = 0 \\ 0 & k_y = \pi \end{cases},
\end{equation}
as can be directly seen from the relation of the model to two mirror-graded copies of the SSH model in the trivial ($k_y=\pi$) and nontrivial ($k_y=0$) phase. 
This confirms that the 2D model is in a topologically nontrivial phase protected by mirror and chiral symmetry. With open boundary conditions, we will therefore find gapless states on both edges with normal to the $x$-direction (see Fig.~\ref{fig: MirrorSSH2D}\textbf{a} for such a geometry), because these are mapped onto themselves under $M_y$. Since the model corresponds to a topological-to-trivial tuning of two copies of the SSH model with opposite winding number, we expect two anti-propagating chiral edge states, which cannot gap out at their crossing at $k_y = 0$ since they belong to different mirror subspaces at this point (see Fig.~\ref{fig: MirrorSSH2D}\textbf{b}). A simple way to see this is that mirror symmetry maps $k_y \rightarrow -k_y$, while it does not change the energy $E$. Therefore it exchanges states pairwise at generic momenta $k_y$ and $-k_y$ and we can form symmetric and anti-symmetric superpositions of them to get mirror eigenstates with eigenvalue $+1$ and $-1$, respectively. The trace of the representation of $M_y$ on this two-dimensional subspace is therefore $0$ at almost all momenta and in particular cannot change discontinuously at $k_y = 0$. Alternatively, direct inspection of the Hamiltonian~\eqref{eq: 2DTCISSHModel} at $k_y=0$ reveals that it is composed of two copies of the SSH model, and in view of the form of the mirror symmetry $M_y$, the two copies reside in opposite mirror subspaces. As a consequence, their end states (the edge modes at $k_y=0$) also have opposite mirror eigenvalues and cannot hybridize.

Another 2D system which has two anti-propagating chiral edge modes is the quantum spin Hall effect protected by time-reversal symmetry, where the edge modes are localized on all boundaries. It corresponds to two Chern insulators, one for spin up and one for spin down. The present model may be viewed as a close relative, where the edge modes are protected by mirror and chiral symmetry as opposed to time-reversal, and are only present on edges preserving the mirror symmetry.

\subsection{Mirror Chern number}
\label{sec: MirrorChern}
In the previous section, we have witnessed an example of a general scheme to construct topological BZ invariants going beyond the tenfold way for systems protected by crystalline symmetries: since a crystalline symmetry acts non-locally in space, it also maps different parts of the BZ onto each other. However, when there are submanifolds of the BZ which are left invariant by the action of the symmetry considered, we may evaluate a non-crystalline invariant on them, suited for the dimension and symmetry class of the corresponding submanifold, as long as we restrict ourselves to one of the symmetry's eigenspaces.

\begin{figure}[t]
\begin{center}
\includegraphics[width=0.75 \textwidth,page=6]{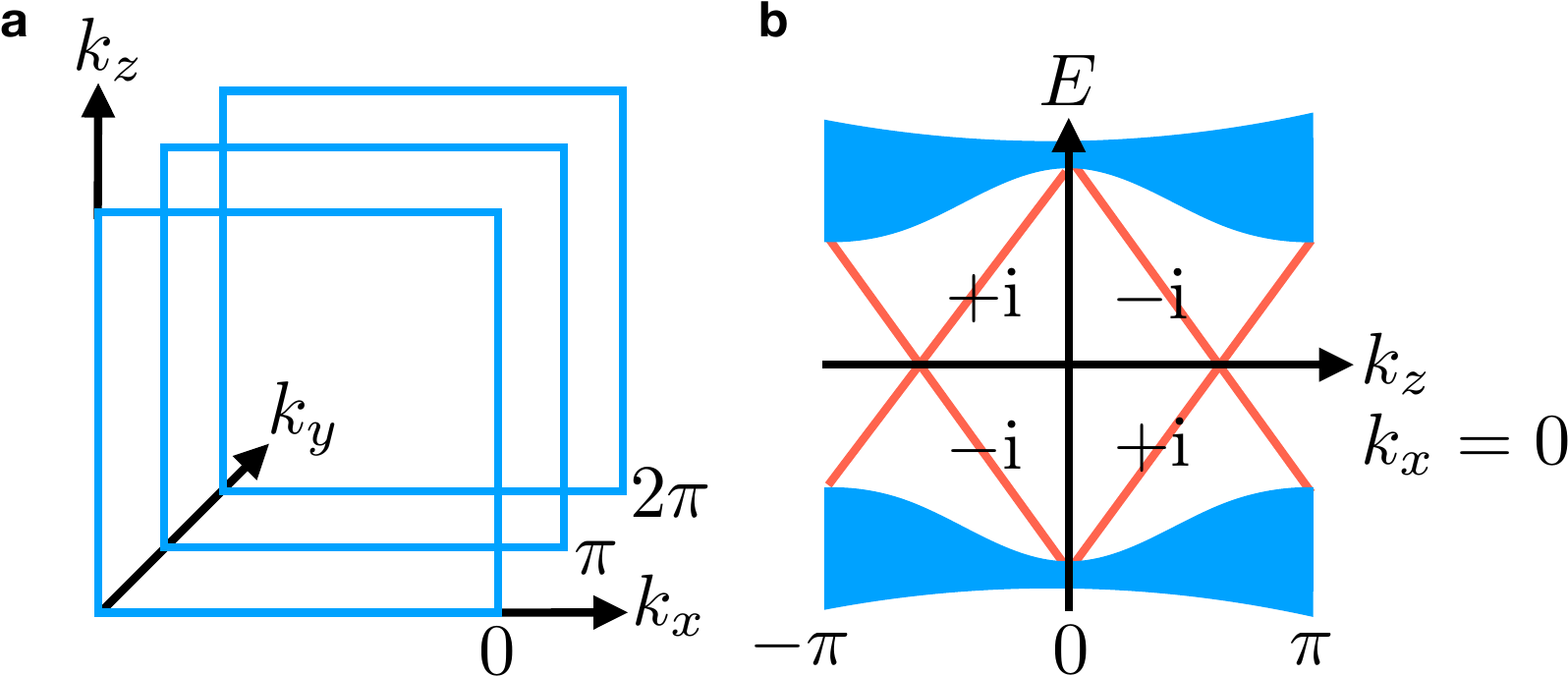}
\caption{
Mirror Chern planes in the BZ and schematic surface spectrum for a time-reversal topological crystalline insulator with $C_\mathrm{m} = 2$. \textbf{a}  For the mirror symmetry $M_y$, there are two planes in the BZ which are left invariant by it and can therefore be used to define a mirror Chern number: the plane at $k_y = 0$ and the one at $k_y = \pi$. \textbf{b}  A mirror Chern number $C_\mathrm{m} = 2$ enforces the presence of two chiral left-movers and two chiral right-movers along mirror symmetric lines in the surface BZ of any surface mapped onto itself by the mirror symmetry. For $M_y$, this is, e.g., the case for the surface obtained by terminating the bulk in $x$-direction and retaining $k_y$ and $k_z$ as momentum quantum numbers. At $k_y=0$ all bands are eigenstates of the mirror symmetry with eigenvalues as shown, and thus prevented from gapping out. At finite $k_y$ however, hybridization becomes possible and we are left with two Dirac cones in the surface BZ in the case at hand.
}
\label{fig: MirrorChern}
\end{center}
\end{figure}

The most prominent example of this construction is the mirror Chern number $C_{\mathrm{m}}$ in three-dimensional systems. Since for a spinful system, mirror symmetry $M$ squares to $M^2 = -1$ its representation in this case has eigenvalues $\pm \mathrm{i}$. Let $\Sigma$ be a surface in the BZ which is left invariant under the action of $M$, such as the surfaces shown in Fig.~\ref{fig: MirrorChern}\textbf{a} for $M_y$. Then, the eigenstates $\ket{u_{\mathbf{k},n}}$ of the Hamiltonian on $\Sigma$ can be decomposed into two groups, $\{\ket{u_{\mathbf{k},l}^+}\}$ and $\{\ket{u_{\mathbf{k},l'}^-}\}$, with mirror eigenvalues $+ \mathrm{i}$ and $- \mathrm{i}$, respectively. Time-reversal symmetry maps one mirror subspace onto the other; if it is present, the two mirror eigenspaces are of the same dimension. We may define the Chern number in each mirror subspace as
\begin{equation}
C_{\pm} = -\frac{\mathrm{i}}{2\pi} \int_\Sigma dk_x dk_z \mathrm{Tr}\left[\mathcal{F}^\pm_{xz}(\mathbf{k})\right].
\end{equation}
Here
\begin{equation}
\mathcal{F}^\pm_{ab}(\mathbf{k})
=
\partial_a\mathcal{A}^+_b(\mathbf{k})-\partial_b\mathcal{A}^+_a(\mathbf{k})
+\left[\mathcal{A}^+_a(\mathbf{k}),\mathcal{A}^+_b(\mathbf{k})\right]
\end{equation}
is the non-Abelian Berry curvature field in the $\pm\mathrm{i}$ mirror subspace, with $\mathcal{A}^\pm_{a;l,l'}(\mathbf{k})=\braket{u^\pm_{\mathbf{k},l}|\partial_a|u^\pm_{\mathbf{k},l'}}$, and matrix multiplication is implied. Since $\mathrm{Tr} \left[\mathcal{A}^+_a(\mathbf{k}),\mathcal{A}^+_b(\mathbf{k})\right] = 0$, this corresponds to Eq.~\eqref{eq: ChernWilsonDef} restricted to a single mirror subspace.
Note that in time-reversal symmetric systems we have $C_{+}=-C_{-}$, and can thus define the mirror Chern number 
\begin{equation}
C_{\mathrm{m}}:=(C_{+}-C_{-})/2.
\end{equation}
A non-vanishing mirror Chern number implies that the Bloch Hamiltonian on $\Sigma$ corresponds to a time-reversal pair of Chern insulators. Thus, the full model will host $C_{\mathrm{m}}$ Kramers pairs of gapless modes on an $M$-invariant line in any surface BZ corresponding to a real space boundary which is mapped onto itself under the mirror symmetry $M$. These Kramers pairs of modes will be generically gapped out away from the lines in the surface BZ which are invariant under the mirror symmetry, and therefore form surface Dirac cones. Indeed, when $C_{\mathrm{m}}$ is odd in a time-reversal symmetric system, this implies an odd number of Dirac cones in any surface BZ, since then the system realizes a conventional time-reversal invariant topological insulator with the Dirac cones located at time-reversal invariant surface momenta. When $C_{\mathrm{m}}$ is even, the surface Dirac cones exist only on mirror symmetric surfaces and are located at generic momenta along the mirror invariant lines of the surface BZ (see Fig.~\ref{fig: MirrorChern}\textbf{b}). This inherently crystalline case is realized in the band structure of tin telluride, SnTe~\cite{Hsieh12}.

\subsection{$C_2 T$-invariant topological crystalline insulator}
Here we present another example of a topological crystalline insulator in 3D, introduced in Ref.~\cite{FuFangTCI2015}, in order to show that surface Dirac cones protected by crystalline symmetries can also appear at generic, low-symmetry, momenta in the surface BZ. We consider a system that is invariant under the combination $C_2 T$ of a two-fold rotation $C_2$ around the $z$ axis and time-reversal symmetry $T$. Note that we take both symmetries to be broken individually.

To understand how this symmetry can protect a topological phase, let us review how time-reversal protects a Dirac cone on the surface of a conventional topological insulator. The effective Hamiltonian on the boundary with surface normal along $z$ of a 3D time-reversal symmetric topological insulator takes the form
\begin{equation}
\mathcal{H} (\mathbf{k}) = k_y \sigma_x - k_x \sigma_y.
\end{equation}
The symmetries are realized as
\begin{eqnarray}
&&T \mathcal{H} (\mathbf{k}) T^{-1} = \mathcal{H} (-\mathbf{k}), \quad T = \mathrm{i} \sigma_y K,\nonumber\\
&&C_2 \mathcal{H} (\mathbf{k}) C_2^{-1} = \mathcal{H} (-\mathbf{k}), \quad C_2 = \sigma_z,
\end{eqnarray}
where we denote by $K$ complex conjugation. Now, the unique mass term for $\mathcal{H} (\mathbf{k})$ which gaps out the Dirac cone is $m \sigma_z$. This term is forbidden by time-reversal as expected, since it does not commute with $T$.

If we dispense with $T$ symmetry and only require invariance under $C_2 T = \sigma_x K$, the mass term is still forbidden. However, the addition of other constant terms to the Hamiltonian is now allowed. The freedom we have is to shift the Dirac cone away from the time-reversal symmetric point $\mathbf{k} = 0$ by changing the Hamiltonian to
\begin{equation}
\mathcal{H} (\mathbf{k}) = (k_y - a) \sigma_x - (k_x - b) \sigma_y,
\end{equation}
with some arbitrary parameters $a$ and $b$. Therefore, the phase stays topologically nontrivial, but has a different boundary spectrum from that of a normal topological insulator. On surfaces preserving  $C_2^z T$ symmetry, any odd number of Dirac cones are stable but are in general shifted away from the time-reversal invariant surface momenta. On the surfaces that are not invariant under $C_2^z T$,  the Dirac cones may be gapped out, since $T$ is broken. This amounts to a $\mathbb{Z}_2$ topological classification of $C_2^z T$-invariant 3D topological crystalline insulators.

\section{Higher-order topological insulators}
So far, when we discussed topological systems in $d$ dimensions, we only considered $(d-1)$ dimensional boundaries which could host gapless states due to the nontrivial topology of the bulk. These systems belong to the class of first-order topological insulators, according to the nomenclature introduced in Ref.~\cite{Schindler17}. In the following, we will give an introduction to second-order topological insulators which have gapless modes on $(d-2)$ dimensional boundaries, that is, on corners in 2D and hinges in 3D, while the boundaries of dimension $(d-1)$ (i.e., the edges of a 2D system and the surfaces of a 3D system) are generically gapped. Higher-order topological insulators require spatial symmetries for their protection and thus constitute an extension of the notion of topological crystalline phases of matter.

\subsection{2D model with corner modes}
\begin{figure}[t]
\begin{center}
\includegraphics[width=0.75 \textwidth,page=7]{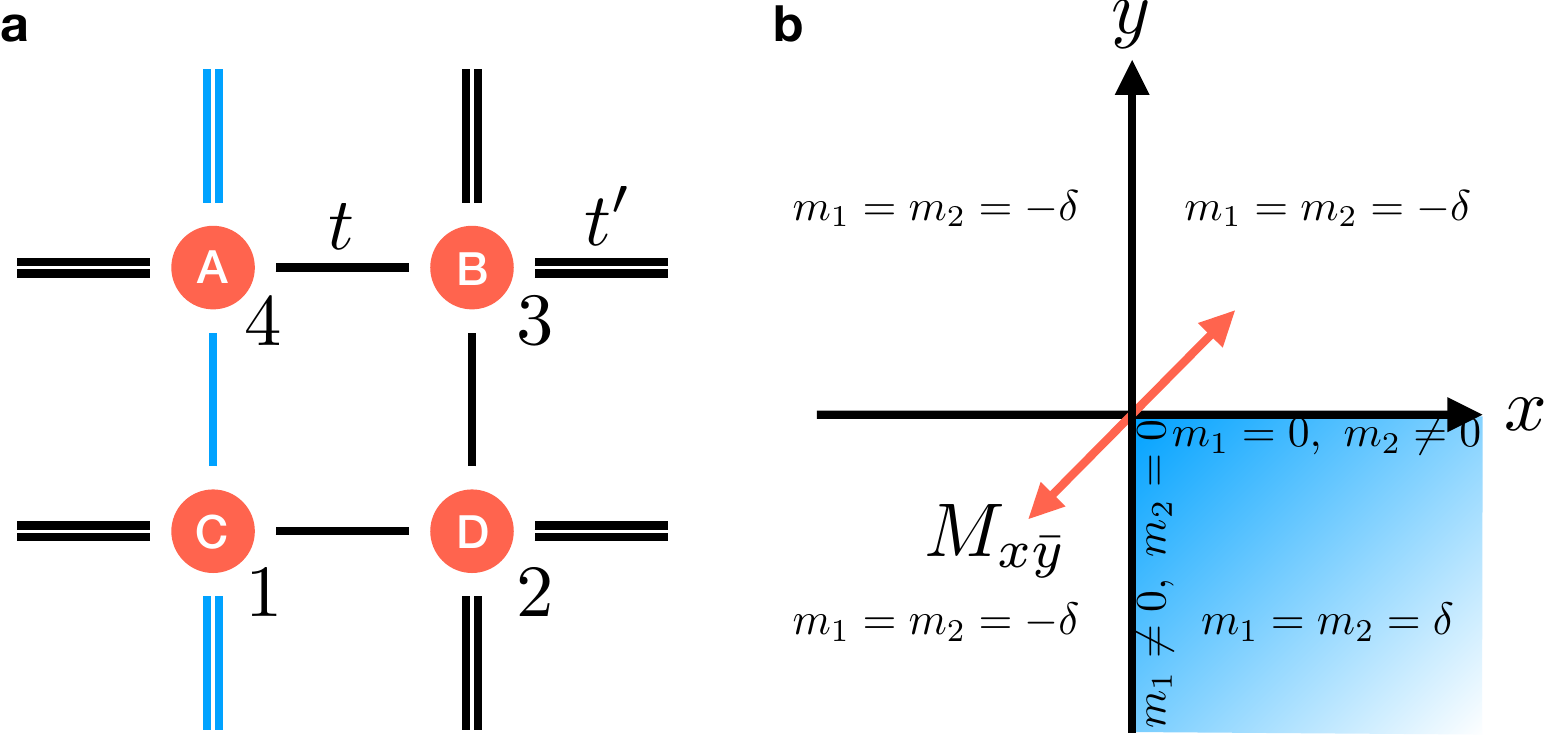}
\caption{
Higher-order 2D SSH model. \textbf{a}  The model features intra-unit cell hopping with strength $t$ as well as inter-unit cell hopping with strength $t'$. For the topological phase we require $t'>t$. In particular, in the fully dimerized limit $t = 0$, $t' \neq 0$ it becomes evident that when we cut the system in two directions to create a corner, there is one dangling site which is not acted upon by any term in the Hamiltonian and therefore provides a zero-mode. The unit cell contains a $\pi$-flux per plaquette, which is realized by  all  blue hoppings being negative, while all black hoppings are positive. \textbf{b}  Vortex geometry of the prefactors of the two masses in Eq.~\eqref{eq: DiracPicture}. We imply a smooth interpolation between the mass values given in the bulk, on the edges, and on the outside. At the corner, the masses vanish and they hence form a vortex-like structure around it.
}
\label{fig: HOTISSH}
\end{center}
\end{figure}

A natural avenue of constructing a higher-order topological phase in 2D is to consider a 2D generalization of the SSH model with unit cell as shown in Fig.~\ref{fig: HOTISSH}\textbf{a} (disregarding the colors in this figure for now) and alternating hoppings $t$ and $t'$ in both the $x$ and $y$-directions. However, naively the bulk of the model defined this way with all hoppings of positive sign is gapless. This can be most easily seen in the fully atomic limit $t'=0$, $t \neq 0$, where the Hamiltonian reduces to a sum over intra-unit cell Hamiltonians of the form
\begin{equation}
\mathcal{H} = t \begin{pmatrix} 0 & 1 & 0 & 1 \\ 1 & 0 & 1 & 0 \\ 0 & 1 & 0 & 1 \\ 1 & 0 & 1 & 0 \end{pmatrix},
\end{equation}
which has obviously zero determinant and therefore gapless modes.

 This was amended in a model introduced in Ref.~\cite{Benalcazar16}, which gave the first example of a higher-order topological insulator, by introducing a magnetic flux of $\pi$ per plaquette. A specific gauge choice realizing this corresponds to reversing the sign of the hoppings along the blue lines in Fig.~\ref{fig: HOTISSH}\textbf{a}. The model then has a gapped bulk, but gapless corner modes. This can be most easily seen in the fully dimerized limit $t=0$, $t' \neq 0$, where one site in each corner unit cell is not acted upon by any term in the Hamiltonian. However, to protect the corner modes we have to include a spatial symmetry in addition to chiral symmetry, since we could otherwise perform an edge manipulation which leaves the bulk (and in particular, its gap) invariant but annihilates one corner mode with another. A natural candidate for this is the pair of diagonal mirror symmetries $M_{xy}$ and $M_{x\bar{y}}$, which each leave a pair of corners invariant and therefore allow for protected gapless modes on them.

Note that we cannot arrive at the same phase by just combining two one-dimensional SSH models glued to the edges of a trivially gapped 2D system: By the mirror symmetry, the two SSH chains on edges that meet in a corner would have to be in the same topological phase. Thus, each would contribute one corner mode. At a single corner, we would therefore have a pair of modes which is not prevented by symmetry from being shifted to finite energies by a perturbation term. This consideration establishes the bulk model we introduced as an intrinsically 2D topological phase of matter. We will now present three alternative approaches to characterize the topology as well as the gapless corner modes of the model.

\subsubsection{Elementary mirror subspace analysis}
The plaquettes along the $x\bar{y}$ diagonal are the only parts of the Hamiltonian mapped onto themselves by the $M_{x\bar{y}}$ mirror symmetry. In the fully dimerized limit $t'\neq0$, $t=0$, we may consider the Hamiltonian as well as the action of $M_{x\bar{y}}$ on a single inter-unit cell plaquette on the diagonal of the system as given by
\begin{equation}
\mathcal{H} =  t' \begin{pmatrix} 0&1&0&-1 \\ 1&0&1&0 \\ 0&1&0&1 \\ -1&0&1&0 \end{pmatrix}, \quad 
M_{x\bar{y}} = \begin{pmatrix} 0&0&1&0 \\ 0&1&0&0 \\ 1&0&0&0 \\ 0&0&0&-1 \end{pmatrix}.
\end{equation}
$M_{x\bar{y}}$ has eigenvectors
\begin{equation}
\ket{+_1} = \begin{pmatrix} 0 \\ 1 \\ 0 \\ 0 \end{pmatrix}, \quad \ket{+_2} = \frac{1}{\sqrt{2}} \begin{pmatrix} 1 \\ 0 \\ 1 \\ 0 \end{pmatrix}, \quad 
\ket{-_1} = \begin{pmatrix} 0 \\ 0 \\ 0 \\ 1 \end{pmatrix}, \quad \ket{-_2} = \frac{1}{\sqrt{2}} \begin{pmatrix} 1 \\ 0 \\ -1 \\ 0 \end{pmatrix}
\end{equation}
with eigenvalues $+1,+1,-1,-1$, respectively. Since $[\mathcal{H},M_{x\bar{y}}] = 0$ we know that the Hamiltonian block-diagonalizes into the two mirror subspaces, and we may calculate its form in each Block separately,
\begin{equation}
\left(\mathcal{H}_+\right)_{ij} = \bra{+_i} \mathcal{H} \ket{+_j} \rightarrow \mathcal{H}_+ = \begin{pmatrix} 0 & \sqrt{2} \\ \sqrt{2} & 0 \end{pmatrix} = \mathcal{H}_-.
\end{equation}
This, however, is exactly the form taken by a single SSH model in the fully dimerized and topologically nontrivial phase (this is because here we have focussed on a plaquette on the diagonal with $t'$ hopping, an adjoining plaquette with $t$ hopping would correspond to the weak bonds in the mirror subspace SSH model). We may therefore interpret our model along one diagonal as two nontrivial SSH models, one for each mirror subspace and protected by the chiral symmetry. Naively this would imply two end modes. However, this is not the case. In the upper left corner, for example, only a single $A$ site is left from an inter-unit cell plaquette (see Fig.~\ref{fig: HOTISSH}\textbf{a}), which happens to have mirror eigenvalue $-1$. Correspondingly, the lower right corner hosts a dangling $D$ site, which has mirror eigenvalue $+1$. Thus, due to this modified bulk-boundary correspondence of the higher-order topological insulator, each of the two diagonal SSH chains has only one end state at opposite ends. These form the corner modes of the higher-order topological insulator.

\subsubsection{Mirror-graded winding number}
In analogy to the mirror graded Wilson loop introduced in Sec.~\ref{sec: 2DTCI}, we can calculate the mirror-graded winding number suited for systems with chiral symmetry. For this we need the full Bloch Hamiltonian, which is given by
\begin{equation}
\label{eq: Hquadrupole}
\mathcal{H}(\mathbf{k}) = (1 + \lambda \cos k_x) \tau_0 \sigma_x + (1 + \lambda \cos k_y) \tau_y \sigma_y - \lambda \sin k_x \tau_z \sigma_y + \lambda \sin k_y \tau_x \sigma_y.
\end{equation}
Note that by a term such as $\sigma_x \tau_0$ we really mean the tensor product $\sigma_x \otimes \tau_0$ of two Pauli matrices.
Here, we have chosen $t = 1$ and $t' = \lambda$. The case where $\lambda > 1$ then corresponds to the topological phase. Along the diagonals of the BZ (and only there), the Hamiltonian may again be block-diagonalized by the mirror symmetries.
Let us consider for concreteness the $\mathbf{k} = (k,k)$ diagonal, which is invariant under  $M_{xy}$ with representation
\begin{equation}
M_{xy}=\begin{pmatrix}
-1&0&0&0\\
0&0&0&1\\
0&0&1&0\\
0&1&0&0
\end{pmatrix}.
\end{equation}
With a transformation that diagonalizes $M_{xy}$, we can bring the Hamiltonian in the form
\begin{equation}
\tilde{\mathcal{H}}(k,k) = \begin{pmatrix}0 & q_+ (k) & 0 & 0 \\ q_+^\dagger (k) & 0 & 0 & 0 \\ 0 & 0 & 0 & q_- (k) \\ 0 & 0 & q_-^\dagger (k) & 0 \end{pmatrix}, \quad q_\pm (k) = \sqrt{2} (1 + \lambda e^{\mp \mathrm{i} k}).
\end{equation}
We see that in the two mirror eigenspaces $\tilde{\mathcal{H}}(k,k)$ takes the form of an SSH model. Defining 
\begin{equation}
\nu_\pm = \frac{\mathrm{i}}{2 \pi} \int \mathrm{d} k \,  \mathrm{Tr}\left[\tilde{q}_\pm (k) \partial_k \tilde{q}^\dagger_\pm (k) \right]
\end{equation}
in analogy to our definition of the one-dimensional winding number, where we have appropriately normalized $\tilde{q}_\pm (k) = q_\pm (k) /|q_\pm (k)|$, we obtain $\nu_\pm = \pm 1$ and therefore $\nu_{M_{xy}} = 1$ for the mirror-graded winding number $\nu_{M_{xy}} = (\nu_+ - \nu_-)/2$. As long as the system obeys the mirror symmetry and the chiral symmetry, $\nu_{M_{xy}}$ is a well-defined topological invariant that cannot be changed without closing the bulk gap of the 2D system.  

\subsubsection{Dirac picture of corner states}
An alternative and very fruitful viewpoint of topological phases of matter arises from the study of continuum Dirac Hamiltonians corresponding to a given phase. For example, the band inversion of a first-order topological insulator can be efficiently captured by the Hamiltonian of a single gapped Dirac cone with mass $m$ in the bulk of the material, and mass $(-m)$ in its exterior. One can then show that the domain wall in $m$ binds exactly one gapless Dirac cone to the surface of the material. We want to develop an analogous understanding of higher-order topological phases as exemplified by the model studied in this section.

For the topological phase transition at $\lambda = 1$ in Eq.~\eqref{eq: Hquadrupole}, there is a gap closing at $\mathbf{k}_0 = (\pi, \pi)$. Expanding $\mathcal{H}(\mathbf{k})$ around this point to first order and setting $\mathbf{k} = \mathbf{k}_0 + \mathbf{p}$, we obtain
\begin{eqnarray}
\label{eq: DiracPicture}
\mathcal{H}(\mathbf{k}) &=& (1 - \lambda) \tau_0 \sigma_x + (1 - \lambda) \tau_y \sigma_y + \lambda p_x \, \tau_z \sigma_y - \lambda p_y \, \tau_x \sigma_y \nonumber\\
&\approx& \delta \tau_0 \sigma_x + \delta \tau_y \sigma_y + p_x \, \tau_z \sigma_y - p_y \, \tau_x \sigma_y,
\end{eqnarray}
where we have defined $\delta = (1-\lambda) \ll 1$ and $\lambda \approx 1$. Note that all matrices anticommute and that there are two mass terms, both proportional to $\delta$, in accordance with the gap-closing phase transition at $\delta = 0$. When terminating the system, a boundary is modeled by a spatial dependence of these masses. We consider the geometry shown in Fig.~\ref{fig: HOTISSH}\textbf{b}, where two edges meet in a corner. The mirror symmetry $M_{x\bar{y}}$ maps one edge to the other but leaves the corner invariant. As a result, the mirror symmetry does not pose any restrictions on the masses on one edge, but once their form is determined on one edge, they are also fixed on the other edge by $M_{x\bar{y}}$. In fact, since $M_{x\bar{y}} \tau_0 \sigma_x M_{x\bar{y}}^{-1} = \tau_y \sigma_y$ with $M_{x\bar{y}} = (\tau_x \sigma_0 + \tau_z \sigma_0 + \tau_x \sigma_z -\tau_z \sigma_z)/2$ and vice versa, we may consider the particularly convenient choice of Fig.~\ref{fig: HOTISSH}\textbf{b} for the mass configuration of the corner geometry. 

From Fig.~\ref{fig: HOTISSH}\textbf{b}, it becomes evident that the symmetries dictate that the masses, when considered as real and imaginary part of a complex number, wind once around the origin of the corresponding complex plane (at which the system becomes gapless) as we go once around the corner in real space. They are mathematically equivalent to a vortex in a $p$-wave superconductor, which is known to bind a single Majorana zero-mode. We can therefore infer the presence of a single gapless corner state for the model considered in this section from its Dirac Hamiltonian.

To be more explicit, denoting by $m_1 (x,y)$ and $m_2 (x,y)$ the position-dependent prefactors of $\sigma_x \tau_0$ and $\sigma_y \tau_y$, respectively, we may adiabatically evolve the Hamiltonian to a form where the mass term vortex is realized in the particularly natural form $m_1(x,y) + \mathrm{i} m_2(x,y) = x + \mathrm{i} y = z$, where $z$ denotes the complex number corresponding to the 2D real space position $(x,y)$. After performing a $C_3$ rotation about the $(111)$-axis in $\tau$ space, which effects the replacement $\tau_x \rightarrow \tau_y \rightarrow \tau_z \rightarrow \tau_x$, and exchanging the order of $\tau$ and $\sigma$ in the tensor product, the resulting matrix takes on the particularly nice form
\begin{eqnarray}
\mathcal{H}(\mathbf{k}) &=& \begin{pmatrix}
0&q(\mathbf{k})\\
q^\dagger(\mathbf{k})&0
\end{pmatrix}, \nonumber\\ 
q(\mathbf{k}) &=& \begin{pmatrix} m_1 - \mathrm{i} m_2 & -\mathrm{i} p_x + p_y \\  -\mathrm{i} p_x - p_y & m_1 + \mathrm{i} m_2 \end{pmatrix} 
= \begin{pmatrix} \bar{z} & -\partial_{\bar{z}} \\ -\partial_{z} & z \end{pmatrix} 
\nonumber\\ \quad \rightarrow \quad q^\dagger(\mathbf{k}) &=& \begin{pmatrix} z & \partial_{\bar{z}} \\  \partial_{z} & \bar{z} \end{pmatrix}.
\end{eqnarray}
While for $q^\dagger(\mathbf{k})$, there is one zero-energy solution $\ket{\Psi} = e^{- z \bar{z}} (1,1)$, the corresponding solution $\ket{\tilde{\Psi}} = e^{\frac{z^2 + \bar{z}^2}{2}} (1,1)$ for $q(\mathbf{k})$ is not normalizable. We thus conclude that there is one zero-mode with eigenfunction $\left(\ket{\Psi},\ket{0}\right)$ localized at the corner of the sample.

\subsection{3D model with hinge modes}

\begin{figure}[t]
\begin{center}
\includegraphics[width=0.75 \textwidth,page=8]{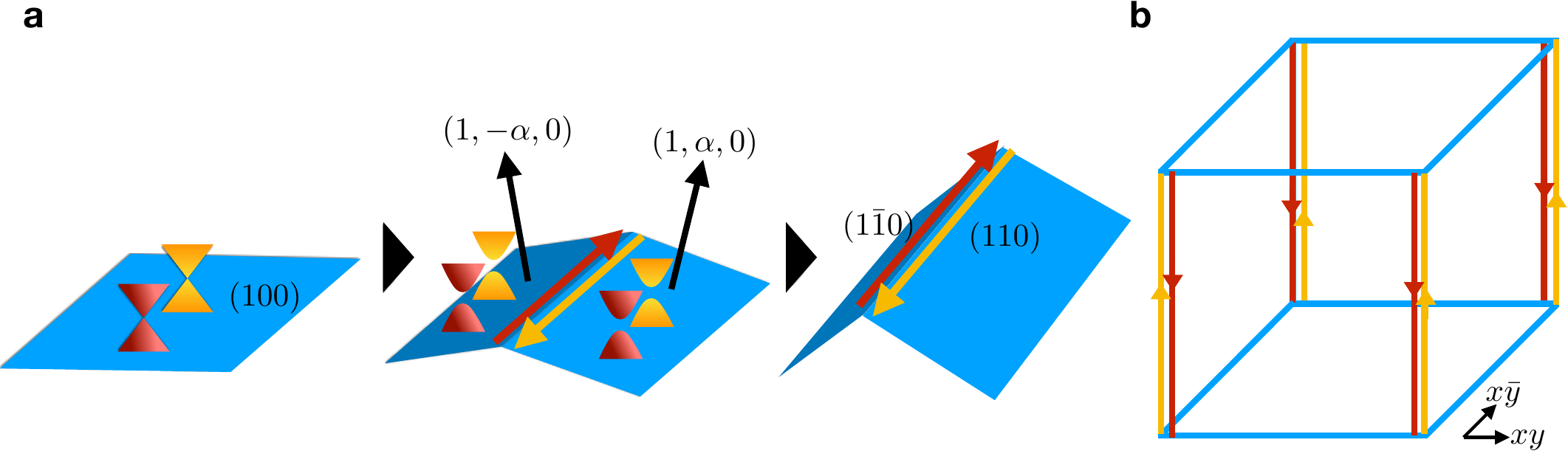}
\caption{
Construction of a 3D second-order topological insulator. \textbf{a} We begin with a surface left invariant by $M_{y}$ on which $C_\mathrm{m} = 2$ implies two gapless Dirac cones. When slightly tilting the surface in opposite directions to form a kink, the Dirac cones on the new surfaces on either side of the kink may be gapped out with opposite masses, since the mirror symmetry maps one into the other and anti commutes with the Dirac mass term. Since a domain wall in a Dirac mass binds a single zero-mode, and the two Dirac cones on each surface are mapped into each other by time-reversal, a Kramers pair of gapless hinge modes emerges on the intersection. When continuing to bend the surfaces to create a right angle, these modes cannot vanish since they are protected by the $M_y$ mirror symmetry. \textbf{b} By this argument we can infer time-reversal paired hinge modes on each hinge along the $x$ (and $y$, if we also take into account the mirror symmetry $M_{x}$ along with $C_\mathrm{m} = 2$) direction.
}
\label{fig: hingeHOTI}
\end{center}
\end{figure}

To construct a higher-order topological insulator in 3D, we start from a time-reversal invariant topological crystalline insulator with mirror Chern numbers in its bulk BZ. For the sake of simplicity we restrict to the case where only the mirror Chern number $C_{\mathrm{m}}$ belonging to the $M_{y}$ symmetry is non-vanishing because the argument goes through for each mirror Chern number separately. We will now show that in an open geometry with surface normals along the $xy$ and $x\bar{y}$ direction, $C_{\mathrm{m}} = 2$ implies the presence of a single time-reversal pair of gapless chiral hinge modes on the intersection of the $(110)$ and $(1\bar{1}0)$ surfaces (see Fig.~\ref{fig: hingeHOTI}).

As discussed in Sec.~\ref{sec: MirrorChern}, for $M_{y}$ symmetry, a nonzero $C_{\mathrm{m}} = 2$ enforces two gapless Dirac cones in the surface BZ of the $(100)$ termination, which is mapped onto itself by $M_{y}$. Note that on this surface, with normal in $x$-direction, $k_y$ and $k_z$ are still good momentum quantum numbers. The Hamiltonian for a single surface Dirac cone can be written as
\begin{equation}
\mathcal{H} (k_y, k_z) = v_1 \sigma_z k_y + v_z \sigma_x (k_z - k_z^0),
\end{equation}
where the mirror symmetry is represented by $M_y = \mathrm{i} \sigma_x$ and thus prevents a mass term of the form $+m \sigma_y$ from appearing.
To arrive at a theory describing the intersection of the $(110)$ and $(1\bar{1}0)$ boundaries, we introduce a mirror-symmetric kink in the $(100)$ surface (see Fig.~\ref{fig: hingeHOTI}\textbf{a}) and so first consider the intersection of two perturbatively small rotations of the $(100)$ surface, one to a $(1,\alpha,0)$ termination and the other to a $(1,-\alpha,0)$ termination with $\alpha \ll 1$. The Hamiltonian on the $(1,\pm \alpha,0)$ surface becomes
\begin{eqnarray}
\mathcal{H}_\pm (k_y, k_z) &=& v_1 \sigma_z (k_y \pm \rho) + v_z \sigma_x (k_z - k_z^0) \pm m \sigma_y \nonumber\\
&\equiv& v_1 \tilde{k}^{\pm}_y \sigma_z + v_z \tilde{k}_z \sigma_x \pm m \sigma_y,
\end{eqnarray}
where $m$ and $\rho$ are small parameters of order $\alpha$ and we have omitted the irrelevant coordinate shifts in the last line. This Hamiltonian describes a gapped Dirac cone, the mass term is now allowed by mirror symmetry since the surfaces considered are no longer invariant under it. Instead, they are mapped onto each other and thus have to carry opposite mass. We note that by this consideration the hinge between the $(1,\alpha,0)$ and $(1,-\alpha,0)$ surfaces constitutes a domain wall in a Dirac mass extended in $z$-direction, which is known to host a single chiral mode~\cite{Jackiw76}.

We will now explicitly solve for this domain wall mode at $\tilde{k}_z = 0$ by going to real space in $y$-direction. Making the replacement $\tilde{k}^{\pm}_y \rightarrow - \mathrm{i} \partial_y$, the Hamiltonian on either side of the hinge becomes
\begin{equation}
\mathcal{H}_\pm = \begin{pmatrix} -\mathrm{i} v_1 \partial_y & \pm \mathrm{i} m \\ \mp \mathrm{i} m & \mathrm{i} v_1 \partial_y\end{pmatrix}.
\end{equation}
$\mathcal{H}_+$, for which $y > 0$, has one normalizable zero-energy solution given by $\ket{\Psi_+} = e^{-\kappa y} (1,1)$ (where we assume $\kappa = m / v_1 > 0$ without loss of generality). $\mathcal{H}_-$, for which $y < 0$, has another normalizable zero-energy solution given by $\ket{\Psi_-} = e^{\kappa y} (1,1)$. Since the spinor $(1,1)$ of the solutions is the same on either side of the hinge, the two solutions can be matched up in a continuous wave function. We obtain a single normalizable zero-energy solution for the full system at $k_z = 0$, which is falling off exponentially away from the hinge with a real-space dependence given by $\ket{\Psi} = e^{- \kappa |y|} (1,1)$. To determine its dispersion, we may calculate the energy shift for an infinitesimal $k_z$ in first-order perturbation theory to find
\begin{equation}
\Delta E(\tilde{k}_z) = \bra{\Psi} v_z \tilde{k}_z \sigma_x \ket{\Psi} = + v_z \tilde{k}_z.
\end{equation}
We have therefore established the presence of a single linearly dispersing chiral mode on the hinge between the $(1,\alpha,0)$ and $(1,-\alpha,0)$ surfaces by considering what happens to a single Dirac cone on the $(1,0,0)$ surface when a kink is introduced. The full model, which by $C_{\mathrm{m}} = 2$ has two $(100)$ surface Dirac cones paired by time-reversal symmetry, therefore hosts a Kramers pair of hinge modes on the intersection between the $(1,\alpha,0)$ and $(1,-\alpha,0)$ surfaces. These surfaces themselves are gapped. 
The two modes forming the hinge Kramers pair have opposite mirror eigenvalue. Increasing $\alpha$ non-perturbatively to $1$ in a mirror-symmetric fashion cannot change the number of these hinge modes, since the chiral modes belong to different mirror subspaces and are thus stable to any perturbation preserving the mirror symmetry.
By this reasoning, we end up with a pair of chiral modes at each hinge in the geometry of Fig.~\ref{fig: hingeHOTI}\textbf{b}.

\subsection{Interacting symmetry protected topological phases with corner modes}
In this last section, we will switch gears and explore how one can construct interacting symmetry-protected topological (SPT) phases of bosons which share the phenomenology of higher-order topological insulators. Note that while non-interacting fermionic systems may have topologically nontrivial ground states, the same is not true for non-interacting bosonic systems whose ground state is a trivial Bose-Einstein condensate \cite{senthil15}. Therefore, for bosons we necessarily need interactions to stabilize a topological phase. We first give a lightning introduction to SPT phases via a very simple model in 1D. A topologically nontrivial SPT state is defined as the gapped ground state of a Hamiltonian, for which there exists no adiabatic interpolation to an atomic limit Hamiltonian without breaking the protecting symmetries or losing the locality of the Hamiltonian along the interpolation~\cite{Chen13}.

\subsubsection{1D model with local symmetry}
Consider a chain of $N$ spin-1/2 degrees of freedom with Hamiltonian
\begin{equation} \label{canonicalH}
H = - \sum_{i=2}^{N-1} A_i, \quad A_i = \sigma_{i-1}^z \sigma_i^x  \sigma_{i+1}^z,
\end{equation}
which describes a system with open boundary conditions.
 All the $A_i$ commute with each other and can therefore be simultaneously diagonalized.

The Hamiltonian $H$ respects a time-reversal $\mathbb{Z}_2^T$ symmetry $[T, H]=0$ represented by the operator 
\begin{equation} \label{Tsym}
T = \mathit{K} \prod_{i} \sigma_i^x.
\end{equation}
Note that $T^2 = +1$.

We now consider a set of operators 
\begin{equation} \label{bigSigmas}
\Sigma^x = \sigma_1^x \sigma_{2}^z, \quad \Sigma^y = \sigma_1^y \sigma_{2}^z, \quad 
\Sigma^z = \sigma^z_1,
\end{equation}
which act locally on the left end of the chain and furnish a Pauli algeba. 
A similar set of operators can be defined for the other end of the chain.

Since $[T, \Sigma^a]_+ = 0$, where $[\cdot,\cdot]_+$ denotes the anti-commutator, these end operators cannot be added as a perturbation to the Hamiltonian without breaking the $\mathbb{Z}_2^T$ symmetry. However, they commute with all the $A_i$ in $H$. This algebra can only be realized on a space with minimum dimension $2$, imposing a twofold degeneracy on the eigenstates of $H$ for each end of the chain. This degeneracy can be interpreted as one gapless spin-1/2 degree of freedom at each end of the chain. Note that a unitary version of $T$ would commute with $\Sigma^y$ rather than anti-commute and therefore not protect these edge degrees of freedom.

\subsubsection{2D model with crystalline symmetry}
We can set up a very similar construction in 2D to arrive at a SPT model with gapless corner modes. Note however that we know from the classification of SPTs by group cohomology~\cite{Chen13} that while in 1D the $\mathbb{Z}_2^T$ symmetry from before indeed protects a $\mathbb{Z}_2$ topological classification, in 2D there is no corresponding nontrivial phase. As was the case for non-interacting fermions, we therefore have to turn to spatial symmetries to protect corner states. Other than that, the construction is very similar to the 1D case.

\begin{figure}[t]
\begin{center}
\includegraphics[width=0.75 \textwidth,page=9]{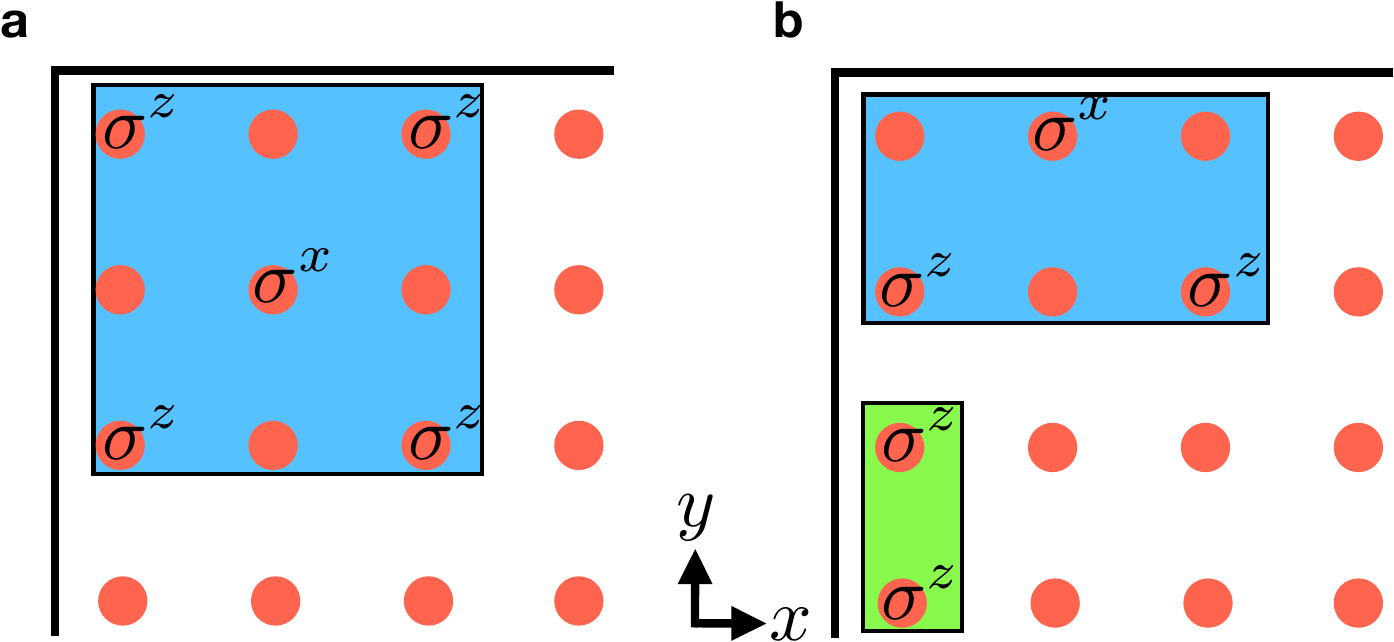}
\caption{
Local operators of a higher-order SPT Hamiltonian with protected corner modes. Each site carries a spin-1/2 degree of freedom acted upon by the Pauli matrices $\sigma^i$, $i=0,x,y,z$. \textbf{a} The bulk Hamiltonian consists of the sum over all sites of tensor products of $\sigma_x$ acting on a given site with $\sigma_z$ on all four adjoining sites along the two diagonals. \textbf{b} Two possible edge Hamiltonian elements which naively would both satisfy the symmetry $M_{x y} T$ when repeated over all edge sites. However, the ground state corresponding to the operator in green spontaneously breaks the symmetry and the operator is thus forbidden. Therefore, we may only terminate the edge with the operator in blue, leading to a two-fold degeneracy in the resulting ground state for each corner.
}
\label{fig: SPTHOTI}
\end{center}
\end{figure}

Consider a square lattice of spin-1/2 degrees of freedom, again with Hamiltonian
\begin{equation}
H = - \sum_{i} A_i, \quad A_i = \sigma_i^x \prod_{j_i \in N(i)} \sigma_{j_i}^z,
\end{equation}
where the set $N(i)$  stands for the four next-to-nearest neighbor sites of site $i$, which are located along the $xy$ and $x\bar{y}$ diagonals (see Fig.~\ref{fig: SPTHOTI}\textbf{a}).
Again, verify that all $A_i$ commute with each other and thus can be simultaneously diagonalized.
We will be interested in open boundary conditions, in which case the sum over $i$ runs only over the interior sites of the lattice, i.e., not the sites on the edges or corners.

Trivially, the model has the same symmetry as given by Eq.~\eqref{Tsym}. However, we need to enrich it with a spatial transformation in order for it to protect topological features. We choose $T = \mathit{K} \prod_i \sigma^x_i$ as before and define
\begin{equation}
M_{x \bar{y}}: \quad (x,y) \rightarrow (-y,-x).
\end{equation}
The model is then invariant under the symmetry $M_{x \bar{y}} T$. We will now show that this symmetry protects a pair of corner states along the $x \bar{y}$ diagonal. In order to also protect states at the other pair of corners we would have to perform the same analysis and require $M_{x y} T$ or $C_4^z$ symmetry in addition.

Indeed, at each corner along the $x \bar{y}$ diagonal, we have a Pauli algebra generated by
\begin{equation}
\Sigma^x = \sigma_c^x \sigma_{j_c}^z, \quad \Sigma^y = \sigma_c^y \sigma_{j_c}^z, \quad \Sigma^z = \sigma^z_c,
\end{equation}
where $c$ denotes the corner site and $j_c$ denotes the site which is the next-to-nearest neighbor of the corner along the diagonal. Crucially, there is a single next-to-nearest neighbor site for each corner, while in the bulk there are four next-to-nearest neighbors along the diagonals.

Since $[M_{x \bar{y}} T, \Sigma^a]_+ = 0$, these corner terms $\Sigma^x$, $\Sigma^y$, and $\Sigma^z$ cannot be added as a perturbation to the Hamiltonian without breaking the symmetry. However, they commute with all the $A_i$ in $H$, again imposing a two-fold degeneracy on the eigenstates of $H$ for each corner. We therefore have one gapless spin-1/2 degree of freedom at each corner lying along the diagonal corresponding to the respective mirror symmetry we require to hold.

Unlike in the 1D case, this is not the end of the story. We have merely shown that each corner provides a two-fold degeneracy, but what about the edge degrees of freedom? In order to arrive at a higher-order phase, we need to gap them out. A natural way to do this is to include in the Hamiltonian not only the $A_i$ terms with four next-to-nearest neighbors, but also the corresponding edge terms which only have two next-to-nearest neighbors. This is in fact symmetry-allowed for all the edge sites except the corners. The terms are sketched in Fig.~\ref{fig: SPTHOTI}\textbf{b}.

We may however also just put an Ising model on the edge by adding the Hamiltonian
\begin{equation}
H_\mathrm{edge} = - \sum_{i \in E} \sigma_i^z \sigma_{N_E(i)}^z,
\label{eq: Edge Ham}
\end{equation}
where $E$ denotes the set of all boundary sites, including the corners (see Fig.~\ref{fig: SPTHOTI}\textbf{b}),
and $N_E(i)$ denotes one of the two nearest-neighbor sites of $i$ on the edge chosen according to an arbitrary but globally fixed edge orientation. Hamiltonian~\eqref{eq: Edge Ham} contains as many terms as there are edge and corner sites combined. The bulk Hamiltonian contains as many terms as there are bulk sites. We want to find a ground state that has simultaneously eigenvalue $+1$ with respect to all these commuting operators. At first sight, because there are as many terms as sites, these constraints fix the ground state completely. However, this is not true. The product of all terms in Hamiltonian~\eqref{eq: Edge Ham} is the identity, because each site is acted upon by two $ \sigma^z$ operators in this product. This means we have globally one less constraint than sites. This degeneracy corresponds to two magnetized ground states of the Ising model that is formed by the gapped edge. Luckily, either of these magnetized ground states of the quantum Ising model in 1D necessarily breaks the $M_{x \bar{y}}$ symmetry (remember that we are, as always in these notes, working at zero temperature). This spontaneous symmetry breaking preempts the definition of our topological case and renders the edge termination defined in Eq.~\eqref{eq: Edge Ham} not permissible.

In conclusion, we have demonstrated how one can construct a 2D higher-order topological phase protected by mirror symmetries, where protection means that the symmetry may not be broken either explicitly or spontaneously in order for there to be gapless corner modes.

\bibliographystyle{spphys}
\bibliography{references}

\end{document}